\documentclass[useAMS,usenatbib,usegraphicx]{mn2e}
\title[Dead Zones and Extrasolar Planetary Properties]{Dead Zones and
Extrasolar Planetary Properties}
\author[Soko Matsumura and Ralph E. Pudritz]{Soko Matsumura$^{1}$\thanks{E-mail:
soko@physics.mcmaster.ca (SM); pudritz@physics.mcmaster.ca (REP)} and 
Ralph E. Pudritz$^{1,2}$\\
$^{1}$Department of Physics and Astronomy, McMaster University, 1280
Main Street West, Hamilton, ON, L8S 4M1, Canada \\
$^{2}$Origins Institute, ABB 241, McMaster University, 1280
Main Street West, Hamilton, ON, L8S 4M1, Canada}
\begin{document}
\date{}
\pagerange{\pageref{firstpage}--\pageref{lastpage}} \pubyear{}
\maketitle
\label{firstpage}
%
%
\begin{abstract}
Most low-mass protostellar disks evolve in clustered environments
where they are affected by external radiation fields, while others
evolve in more isolated star-forming regions.
Assuming that the magneto-rotational instability (MRI) is the main
source of viscosity, we calculate the size of a poorly ionized, MRI
inactive, and hence low viscosity region -- the ``dead zone'' -- in these
protostellar disks.
We include disk ionization by X-rays, cosmic rays, radioactive
elements and thermal collisions, recombination by molecules, metals,
and grains, as well as the effect of turbulence stimulation in the
dead zone by the active layers lying above it.
We also calculate the gap-opening masses of planets, which are
determined by a disk's viscosity and a disk aspect ratio, for disks in
these environments and compare them with each other.

We find that the dead zone is a robust feature of the protostellar
disks that is largely independent of their environment, typically
stretching out to \(\sim 15\) AU.
We analyze the possible effects of dead zones on planet formation,
migration, and eccentricity evolution.
We show that the gap-opening mass inside the dead zone is
expected to be of the order of terrestrial and ice giant mass planets
while that outside the dead zone is Jovian or super-Jovian mass
planets, largely independent of the star-forming environment.
We show that dead zones can significantly slow down both type I
and type II planetary migration due to their lower viscosity.
We also find that the growth of eccentricity of massive extrasolar planets
is particularly favorable through the planet-disk interaction inside
the dead zones due to the large gaps expected to be opened by planets.
\end{abstract}
\begin{keywords}
accretion, accretion disks - turbulence - planetary systems: formation
- planetary systems: protoplanetary disks - planets and satellites:
general - Solar system: formation - stars: pre-main-sequence
\end{keywords}
%
%
\section{Introduction}
The discovery of over 160 extrasolar planets has opened up several
fundamental questions about planet formation.
Planetary masses vary from \(7.5\) Earth masses all the way up to \(\sim
15\) Jupiter masses.
Massive planets are found close to their central stars which implies
that their migration from their points of origin further out in the
disk needs to be understood.
The high eccentricity of many of the planets also demands explanation.
Theoretical attention has increasingly focused on the interaction of
planets and their surrounding disks in order to elucidate these
questions.
In this regard, while most models still assume that disks possess a
constant turbulent viscosity throughout, the possibility that ``dead
zones'' of very low turbulence exist in disks has profound effects on
planet formation \citep[][hereafter MP03]{Matsumura03}.

The idea of a dead zone in protostellar disks was first proposed by \cite{Gammie96}.
He argued that there is a poorly ionized region where the growth of
the magneto-rotational instability (MRI) against Ohmic dissipation
cannot be sustained.
Since the MRI is thought to be the most promising source of the disk's
turbulent ``viscosity'', the dead zone is expected to have nonzero,
but very low viscosity.

Dead zones have several important effects on planet formation including
(1) planetary masses, (2) planetary migration, (3) planet formation
via gravitational instability, and (4) planetary eccentricity.
We discuss these below. 

The first point was analyzed in our previous paper
\citep[][hereafter MP05]{Matsumura05}: the gap-opening mass of planets
is expected to be smaller inside the dead zone.
A planet opens a gap when the angular momentum transfer rate by the
planetary tidal torque exceeds that of the disk's viscous torque
\citep[e.g.][]{Lin93}. 
Since a dead zone has a very low viscosity, the gap-clearing timescale
is sufficiently short compared to the gas-accretion timescale in that
region so that a planet won't accrete much gas. 

Regarding the second point, dead zones are likely to prolong the planetary
migration timescale \citep[e.g.][MP05]{Chiang02}.
This is potentially important because the migration timescale estimated
in a protostellar disk without a dead zone tends to be much shorter
than a disk's lifetime, and as yet we have no obvious mechanism to
stop planet migration.
Protoplanets migrate as a result of their resonant interaction with
protostellar disks. 
They tend to migrate inward because the magnitude of the outer torque
is usually larger than the inner torque \citep{Ward97}.
The current theories suggest that planetary migration has essentially
two stages.
The first stage is when protoplanets are not massive enough to open a
gap in the disk; they migrate through the disk (type I) as they
accumulate mass.
The second stage starts when protoplanets become sufficiently massive
to open a gap; they migrate at the viscous timescale of the disk (type
II) without accumulating much mass.
A dead zone can affect both types of migration.
A dead zone's effect on type II migration can be readily understood --
a planet's migration speed is significantly slowed down as soon as they
enter a dead zone.
This is because the speed of type II migration is directly
proportional to a disk's viscosity and because the viscosity in a dead
zone is expected to be very low.

A dead zone's effect on type I migration is rather indirect -- planet
migration can be halted or even reversed by the mass accumulation
at the edge of the dead zone \citep[e.g.][]{Thommes05}. 
Even when planets are not stopped at the edge of the dead zone, their
migration speed is still likely to be reduced as soon as they enter it.
This is because these planets will probably open gaps in the dead
zone where the gap-opening mass is roughly a couple of orders of
magnitude smaller.
These planets then switch to the type II migration that is a few orders
of magnitude slower than the type I migration \citep{Ward97}.
Thus, the dead zone may rescue planets from plunging into the central
star due to its lower viscosity, and may also work as a switch from
type I to type II migrations or even as a wall to halt or reverse the
type I migration.
We will further discuss this point in \S5.

Regarding the third point, the mass accumulation at the edge of a dead
zone, or even at the edge of a gap opened by a planet, may lead
to planet formation via gravitational instability, which requires
a dense, cold region \citep[e.g.][]{Boss97,Mayer02}. 
The former type of mass accumulation occurs due to different evolution
speeds between dead zones and active zones, while the latter mass
accumulation happens because the gap-opening timescale tends to be
faster than the disk's viscous evolution timescale.
\cite{Lufkin04} showed that higher density spiral arms formed by a
massive planet can lead to a subsequent planet formation by
gravitational instability.

Regarding the fourth and final point, we propose in this paper
that the dead zone may be an optimum place for the eccentricity
evolution, especially for a disk-planet interaction scenario proposed
by \cite{Goldreich03} .
The lower viscosity inside the dead zone would make a wider gap around
a planet, and therefore allow more space and time for eccentricity growth.
We discuss this point in \S6.

In this paper, we generalize our work in MP05 and calculate the size
of dead zones and planetary gap-opening masses in regions of clustered
and isolated star-forming environments.
This allows us to examine how much the size of dead zones is changed
by its environment, and therefore which parameters affect them most. 
We also discuss each possible effect of a dead zone proposed above.
We find that both the size of a dead zone (typically \(\sim 10-20\)
AU) and gap-opening masses are almost independent of a disk's
environment. 
This is because the disk structures and hence the ionization
structures in dense regions of disks (e.g. a possible region of a
dead zone) are almost the same for both isolated and clustered
star-forming environments.
We give an overview of a gap-opening mass in \S2, and the MRI
turbulence and our disk models in \S3.
We present our results in \S4 and discuss dead zone effects on
migration and eccentricity growth in \S5 and \S6 respectively.
Finally, we summarize our work in \S7.
\section{Gap-opening Mass}
The mechanisms of angular momentum transfer in protostellar disks are
not fully understood, but planet formation is thought to be linked to
at least two of them -- tidal interaction between a gaseous disk and a
protoplanet, and viscous diffusion of a disk.

Tidal torques carry away angular momentum most efficiently around
Lindblad resonances located at about a disk pressure scale height
\(h\) from a protoplanet \citep{Goldreich80} and transfer that to the
disk when generated density waves shock and damp.
Assuming the disk is inviscid so that there is no viscous torque, and
that the density waves shock immediately in the vicinity of the
Lindblad resonances, \cite{Lin93} determined a gap-opening mass of the planet:
\begin{equation}
\label{go1}
\frac{M_{p}}{M_{*}} \geq 3\left(\frac{h_{p}}{r_{p}}\right)^3 \ ,
\end{equation}
where \(M_{p}\) and \(M_{*}\) are mass of a planet and a star
respectively, \(r_{p}\) is the orbital radius of the planet, and
\(h_{p}\) is the pressure scale height at that radius.
\cite{Rafikov02} argued that their assumption is too radical and
determined the gap-opening mass in an inviscid disk by considering the
nonlinear evolution of density waves as well as taking account of the
planet migration effect:
\begin{equation}
\label{go2}
\frac{M_{p}}{M_{*}} > \frac{2}{3}\left(\frac{h_{p}}{r_{p}}\right)^3
{\rm min}\left[5.2 Q^{-5/7}, 3.8 \left(Q \frac{r_{p}}{h_{p}}\right)^{-5/13}\right] \ ,
\end{equation}
where \(Q\) is the Toomre parameter.
The first term represents the case of feedback being not strong enough to
stop migration -- the density waves reflected from the edge of a
forming gap are not enough to fill in the gap and hence smooth out the
difference between outer and inner planetary torques. 
The second term corresponds to the case of feedback being sufficient
to stop migration.
All the disk models we used turn out to be gravitationally stable (\(Q\gg1\)),
thus planets are expected to form through the core accretion rather than
the gravitational instability, and hence have smaller gap-opening
masses compared to Equation (\ref{go1}). 
Equation (\ref{go1}) replaces Equation (\ref{go2}) when density
waves damp immediately at the Lindblad resonances (at a distance of
\(\sim h_{p}\) from the planet) where they are excited.  

When the disk is viscous, we need to take account of the effect of
viscous torque whose primary source is thought to be the MRI turbulence.
\cite{Balbus91} showed that a slightly perturbed, weak magnetic field
can be amplified sufficiently to radially transfer the angular momentum.
Recent numerical simulations \citep[e.g.][]{Fleming00} have shown that
the MRI is active when the dissipation timescale of the turbulence
(\(t_{dis}=h^2/\eta\)) is more than \(10^3 - 10^4\) times slower than
the growth timescale (\(t_{growth}=h/V_A\)):
\begin{equation}
\label{rem}
Re_{M}=\frac{V_{A}h}{\eta} > 10^3 - 10^4 \ ,
\end{equation}
where \(V_{A}\) is the Alfv\'{e}n speed and \(\eta\) is the
diffusivity of the magnetic field.
\footnote[1]{\cite{Sano02} (hereafter SS02) studied the nonlinear evolution
of the MRI in weakly ionized accretion disks including the Hall
effect, and obtained the critical magnetic Reynolds number of
\(Re_{M,SS02}=V_{A}^2/(\eta \Omega)\sim30\).  Our definition of
the magnetic Reynolds number relates to theirs as
\(Re_{M,SS02}=V_{A}/c_{s} \ Re_{M}=\alpha_{ss}^{1/2}Re_{M}\), where
\(c_{s}\) is the sound speed. 
Therefore, their critical magnetic Reynolds number roughly corresponds
to the case of \(Re_{M}=10^2\) for the standard viscosity parameter \(\alpha_{ss}=0.01\).}   

In a viscous disk, the gap-opening mass is reached when the angular
momentum transfer rate by the tidal torque exceeds that by the viscous
torque \citep[e.g.][]{Lin93}:
\begin{equation}
\label{go3}
\frac{M_{p}}{M_{*}} \geq \sqrt{40 \alpha_{ss}
\left(\frac{h_{p}}{r_{p}}\right)^5} \ ,
\end{equation}
where \(\alpha_{ss}\) is the viscous \(\alpha\) parameter \citep{Shakura73}.
For a less viscous disk, or equivalently, a disk with a smaller
\(\alpha_{ss}\), a planetary gap-opening mass becomes smaller.
Assuming that the gap-opening mass inside the dead zone is well
approximated by the gap-opening equation (\ref{go2}), MP05
showed that the gap-opening mass ratio inside the dead zone to outside
it could be up to about 100 -- close to the mass ratio of Jovian to
terrestrial planets in our solar system.

Although numerical simulations show that there is some residual gas
accretion through a gap \citep[e.g.][]{Kley99,Lubow99}, the
gap-opening mass is still expected to be close to a planet's final mass.
\cite{Kley99} showed that the gas accretion through a gap is markedly
reduced for a small viscosity: \(\alpha_{ss}\le5\times10^{-4}\).
This corresponds to a viscosity inside a dead zone found in numerical
simulations \citep[e.g.][]{Fleming03}.
Thus, we can expect that the gap-opening mass is most likely to be the final
planetary mass inside a dead zone.
In a standard disk with mass of \(\sim 0.02 M_{\odot}\),
\cite{Lubow99} showed that the gas accretion rate through a gap is
\(10^{-5}-10^{-6} M_{J}/{\rm year}\) at \(10-40\) AU -- it takes
\(10^5-10^6\) years to double a mass of Jupiter outside a dead zone.
This may be shorter than the disk lifetime, but longer than the
planetary migration timescale in this region -- type I migration
timescale for Jupiter is \(10^3-10^4\) years and type II migration
timescale for \(\alpha_{ss}=0.01\) is \(10^4-10^5\) years (see
Fig. \ref{fig10}).
Thus, as long as planets migrate inward, they probably enter a dead
zone and thereby stop gas accretion (as they open a clear gap) before
they acquire a significant amount of mass through gaps.
\section{MRI and Disk Models}
\subsection{MRI -- its existence and effectiveness}
Although the MRI is considered to be an effective potential source of
disk viscosity, its existence and effectiveness are still under discussion.
Regarding the former point, \cite{Fromang02} pointed out that a very
small fraction of the cosmic abundance of metal atoms can
significantly diminish the size of a dead zone by picking up the
charges of molecular ions which recombine with electrons \(10^5\)
times more effectively than metal ions (i.e. keeping the overall
ionization rate high).
The electron fraction, however, can be significantly reduced by grains
\citep[e.g.][]{Sano00} because they recombine with electrons even more
efficiently than molecular ions.
This effect can potentially make the midplane region of an entire
protostellar disk MRI dead.
We consider recombination rates of electrons with metal ions,
molecular ions as well as grains as in MP05 (also see Appendix A in
this paper).
To calculate the ionization rates, we take account of X-rays, cosmic rays,
radioactive elements and thermal collisions of alkali ions (for
details, see MP03 \& MP05).

Dead zones have been studied by many authors both analytically
\citep[e.g.][MP03, MP05]{Gammie96,Sano00,Glassgold00,Fromang02} and
numerically \citep[e.g.][]{Fleming03}.
Recently, \cite{Inutsuka05ap} suggested that once MRI turbulence
becomes active, it can be sustained in a protostellar disk without any
external ionization and concluded that most regions in protostellar
disks remain magnetically active. 
For dense regions of a disk with dust grains, they proposed that
collisions of energetic electrons might provide enough ionization to
sustain MRI.
In considering their arguments however, we note that their estimate
for the electron fraction is still one to two orders of magnitude
smaller than the one required by recent numerical work.
For a disk surface, assuming dust grains are depleted from the region,
they proposed that turbulent eddies could mix the ionized region with
the neutral region, and therefore homogenize the ionization region, if
the recombination rate is sufficiently low.
Grains could indeed be absent in a high temperature region close to
the central star (see \S4.3).
However, both their effective electron recombination
coefficient and the initial electron fraction are one to two orders of
magnitude too small because of their assumption that \(\alpha_{ss}=0.2\).
\footnote[2]{Using the formula given in \cite{Inutsuka05ap}, the
effective recombination coefficient can be written as \(\beta^{\prime}
\sim 10^{-7}\frac{n_{m^+}}{n_{e}} + 10^{-12}\frac{n_{M^+}}{n_{e}}\),
where \(n_{e}\), \(n_{m^+}\), and \(n_{M^+}\) are number densities of
electrons, molecular ions, and metal ions respectively (see Appendix A).
At regions close to the central star, densities of electrons and metal
ions are about two orders of magnitude larger than the density of
molecular ions \citep{Sano00}.  This will give
\(\beta^{\prime}\sim10^{-9}\) and the electron fraction of
\(\sim10^{-14}\) \citep[see eq. (12) in][]{Inutsuka05ap}, unless
the initial electron fraction is larger.  Also, their suggested
electron fraction to sustain MRI is likely an underestimate because
they adopted a relatively large viscosity parameter \(\alpha_{ss}=0.2\).}
Thus, we consider these processes are probably unable to sustain the MRI
turbulence throughout the entire disk.

Even if the MRI is present, other effects like ambipolar diffusion
could work against its angular momentum transport.
Using three-dimensional magneto-hydrodynamic simulations,
\cite{Hawley98} showed that there is significant angular momentum
transfer when the ion-neutral collision rate is 100 times larger than
the local epicyclic frequency, and that ions and neutrals are essentially
decoupled when this ratio goes below 0.01.
We calculated the corresponding ratio in our disk models and found
that the value is typically \(\sim 10\).
Therefore, we expect at least some angular momentum transport in
active regions of our disk models.

\cite{Fleming03} studied the evolution of the MRI in stratified
accretion disks and found that there is a minimum level of angular
momentum transport in the dead zone in the presence of active layers
due to the Reynolds stress.
In the case relevant to ours (\(\Sigma_{DZ}/\Sigma_{ALs} \sim 10\)),
they obtained \(\alpha_{ss} \sim 10^{-3}\) for active layers and
\(\alpha_{ss} \sim 5 \times 10^{-5}\) for a dead zone.
Later we will show that this \(\alpha_{ss}\) value doesn't affect the
gap-opening mass inside the dead zone.
Another effect discussed in the literature is the mass mixing between
a dead zone and the active layers lying above it \citep{Fleming03}.
Although this vertical mixing may have no effect in radial angular
momentum transfer, we take account of this effect by imposing the
condition \(\Sigma_{DZ}/\Sigma_{ALs} \geq 10\) on a dead zone besides 
\(Re_{M} < 10^3 - 10^4\).
\subsection{Disk Models -- Clustered vs Isolated Star Formation}
Typical protostellar systems are expected to form in a cluster like
Orion Nebula, where they are irradiated by nearby luminous OB stars.
On the other hand, some star formation regions like Taurus are known not
to have young massive stars \citep[e.g.][]{Luhman03}.
It is interesting to compare these two star formation regions from the
viewpoint of their planet formation environments.
 
To determine the size of dead zones of disks in these environments, we
need to calculate \(Re_{M}\) (see equation (\ref{rem})) throughout
disks that depends both on the thermal structure of the disks via \(h\)
and on the ionization structure via \(\eta\).
Both of these quantities will be affected by the presence of massive stars.

\begin{figure}
\unitlength1cm
\begin{minipage}[t]{18.5cm}
\hspace{0.5cm}
\begin{minipage}[h]{14cm}
\begin{picture}(5.5,6)
\scalebox{0.28}{
\includegraphics{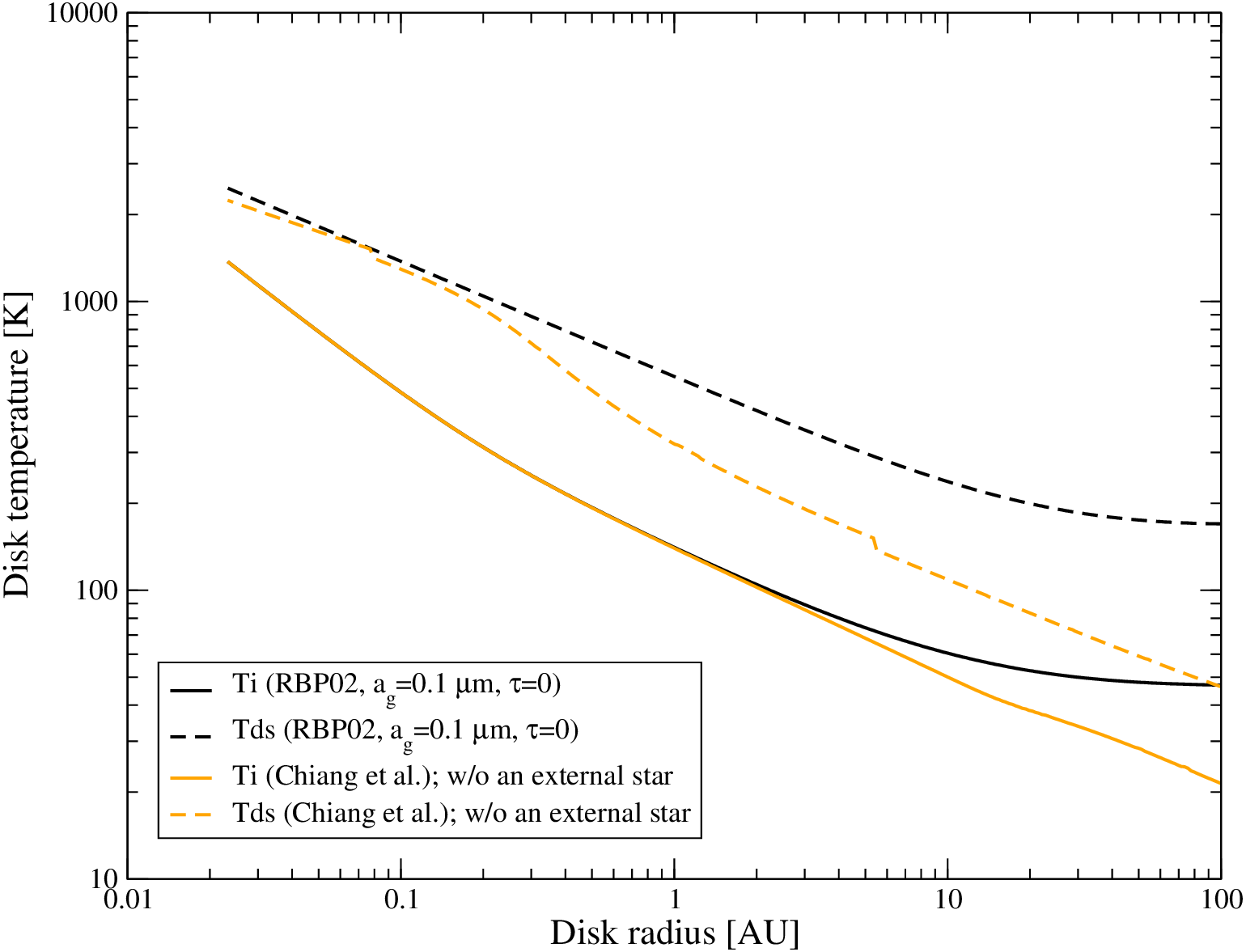}}
\end{picture}
\hspace{2cm}
\begin{picture}(5.5,6)
\scalebox{0.28}{
\includegraphics{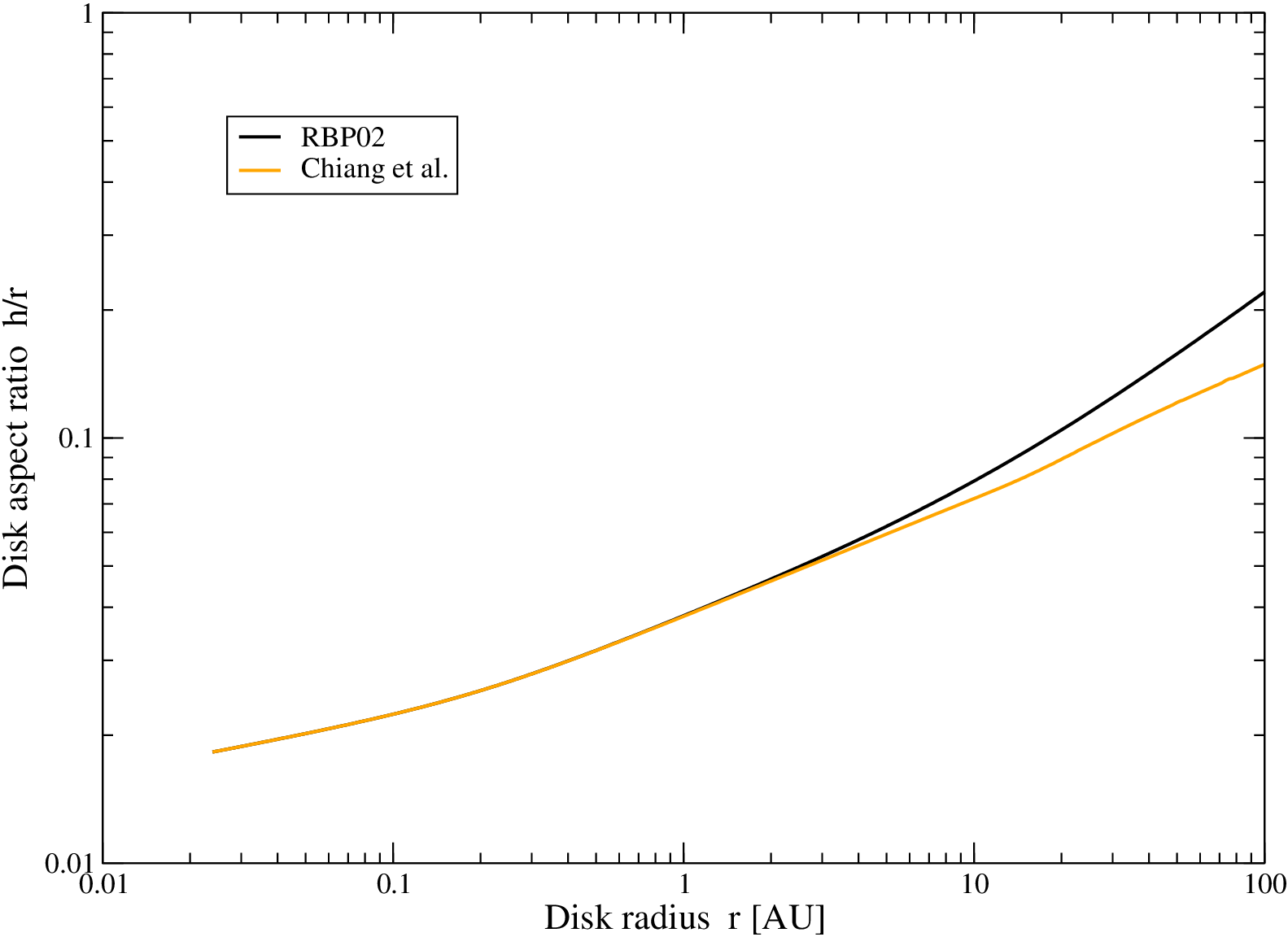}}
\end{picture}
\end{minipage}
\end{minipage}
\begin{minipage}[t]{18cm}
\vspace{0.5cm}
\caption[temp]{Left: temperature structures of disk models in
clustered (heavy lines) and isolated (light lines) environments.  Solid
lines show disk temperatures below the pressure scale height of the
disk, and dashed lines show disk envelope's temperatures.  Right: disk
aspect ratios of disk models in clustered (heavy line) and isolated
(light line) environments.  External fields' effects become important
in an outer part of the disks. \label{fig1}}
\end{minipage}
\end{figure}

The difference in the thermal structure for protostellar disks in the
clustered and isolated case is that the former disks are exposed not
only to their central stars, but also to other massive stars in the cluster.
For the isolated disks, as in MP03, we adopted the isolated,
two-layered disk models developed by \cite{Chiang01}.
For the disks in a cluster, as in MP05, we used the disk models by
\cite{Robberto02} where they modified the isolated two-layered disk
models of \cite{CG97} by submerging these disks in a cluster
environment.
In their model, the external star is assumed to be located at an
average distance of 0.1 pc from a low-mass protostellar disk and have
a stellar luminosity of \(L_{*}=6 \times 10^{38} \ {\rm ergs \ s^{-1}}\).
Due to this extra heating, the disks in a cluster are more flared
compared to the isolated cases (see Fig. \ref{fig1}).
In both disk models, disk temperatures are calculated by assuming a
radiative, hydrostatic equilibrium.
For a particular disk radius, these models give two temperatures -- one in
the surface layer, and the other in the disk interior.
Also, both disk models assume a power-law surface mass density:
\(\Sigma=\Sigma_{0}\left(\frac{r}{{\rm AU}}\right)^{-3/2}\).

Fig. \ref{fig1} compares temperatures of these models (left panel) as well as
corresponding disk aspect ratios (right panel).
From the left panel, it is apparent that the disk interior temperatures
(i.e. the mid-plane temperatures) are the same for both environments
out to \(\sim 10\) AU, while the radiation from external fields
clearly affect on a disk's thermal structure beyond \(\sim 10\) AU,
giving a nearly constant temperature (\(\sim 50\) K) there.
The temperature difference of up to a factor of \(\sim 2\) leads to a
more flared disk, and hence a larger disk aspect ratio and larger
gap-opening masses in a clustered star-forming environment
(see Equation (\ref{go2}) and (\ref{go3})).   
From the right panel, we can see that, at 20 AU, the disk aspect ratio
for isolated and clustered star-forming regions are 0.09 and 0.15 respectively.
This \(\sim 70\) \% difference in a disk aspect ratio leads to a
gap-opening mass of \(0.8\) and \(1.2 M_{J}\) (where \(M_{J}\) is a
mass of Jupiter) in each environment for a standard alpha parameter of
\(\alpha_{ss}=0.01\) (see also: Fig. \ref{fig7}).
Note that, by knowing the radial temperature structure of the disk,
and hence its aspect ratio; \(h/r\), the only parameter needed to
determine the distribution of gap-opening masses is a disk's viscosity
parameter \(\alpha_{ss}\).
We will discuss this in \S 4.1 and 4.2.
Also, the temperature at the innermost disk reaches \(\sim 1500\) K,
suggesting that the region may be void of dust grains due to thermal
evaporation.
We will consider this effect in \S 4.3.

Note that, although the estimated surface layer's temperature of the
disk model in a clustered environment is very low; \(\sim 200\) K
beyond 10 AU, compared to the atmosphere temperature assumed for
photoevaporating disk models; \(\sim 10^4\) K
\citep[e.g.][]{Shu93,Hollenbach94}, it does not affect our results so much.
This is because the disk atmosphere reaches that high temperature only
in very high altitude regions -- about one order of magnitude higher
than our surface disk height, where the density is very low.
In addition, the gap-opening masses depend on the disk interior
temperature, which is related to the pressure scale height rather than
the disk surface temperature.

The difference in the ionization structure for protostellar disks in
the clustered and isolated case is that the former is also exposed to the X-ray
radiation from an external star besides other ionization sources
common in both cases (see below).
Here, we assume the X-ray luminosity of an external star to be
\(L_{x}=10^{34} \ {\rm ergs \ s^{-1}}\), which is likely to be an
upper limit \citep{Stelzer05ap}.
\footnote[3]{We performed the simulation with a more typical value of
the X-ray luminosity of a massive star: \(L_{x}=10^{31} \ {\rm ergs
\ s^{-1}}\), and found that the external star has less effect on the
total ionization rate.  But this does not change the extent of the
dead zones.}
In both environments, the ionization of disks is due to X-rays from
the central star (\(L_{x}=10^{30} \ {\rm ergs \ s^{-1}}\) with
\(kT_{x}=2\) keV, \cite{Feigelson02}), cosmic rays (\(\xi_{CR}\sim
10^{-17} \ {\rm s^{-1}}\) with the attenuation length of \(\sim 96 \
{\rm g \ cm^{-2}}\), \cite{Sano00}), radioactive elements (\(\xi_{RA}\sim
6.9 \times 10^{-23} \ {\rm s^{-1}}\), \cite{Umebayashi81}), as well as
heated alkali ions (important in the high temperature, inner part of
the disks).
\section{Dead Zones and Gap-opening Masses}
In this section, we compare protostellar disks in different
environments and obtain sizes of their dead zones (\S
4.1) as well as the gap-opening masses of planets (\S 4.2).
Our focuses are on comparing (1) disks in a stellar cluster and
isolated disks, and (2) disks with and without cosmic ray ionization.
The latter case is included because low energy cosmic rays, which are
more responsible for ionization compared to higher energy cosmic
rays \citep[e.g.][]{Lepp92}, may be partially excluded from the disks
by magnetic scattering \citep{Skilling76}.
Note, however, that \cite{Desch04} recently showed that galactic
cosmic ray radiation is likely to be abundant in protostellar disks.
Here we include this comparison for completeness.
In the end of this section, we also discuss the effect of dust
evaporation (\S 4.3).
\subsection{Clustered vs Isolated Star Formation -- Dead Zones}
\begin{figure}
\unitlength1cm
\begin{minipage}[t]{18.5cm}
\hspace{3.5cm}
\parbox[t]{5.5cm}{{Disks in a cluster}}
\hspace{2cm}
\parbox[t]{5.5cm}{{Isolated Disks}}
\end{minipage}
\begin{minipage}[t]{18.5cm}
\hspace{0.5cm}
\begin{minipage}[h]{14cm}
\mbox{} \\
\begin{picture}(5.5,6)
\scalebox{0.28}{
\includegraphics{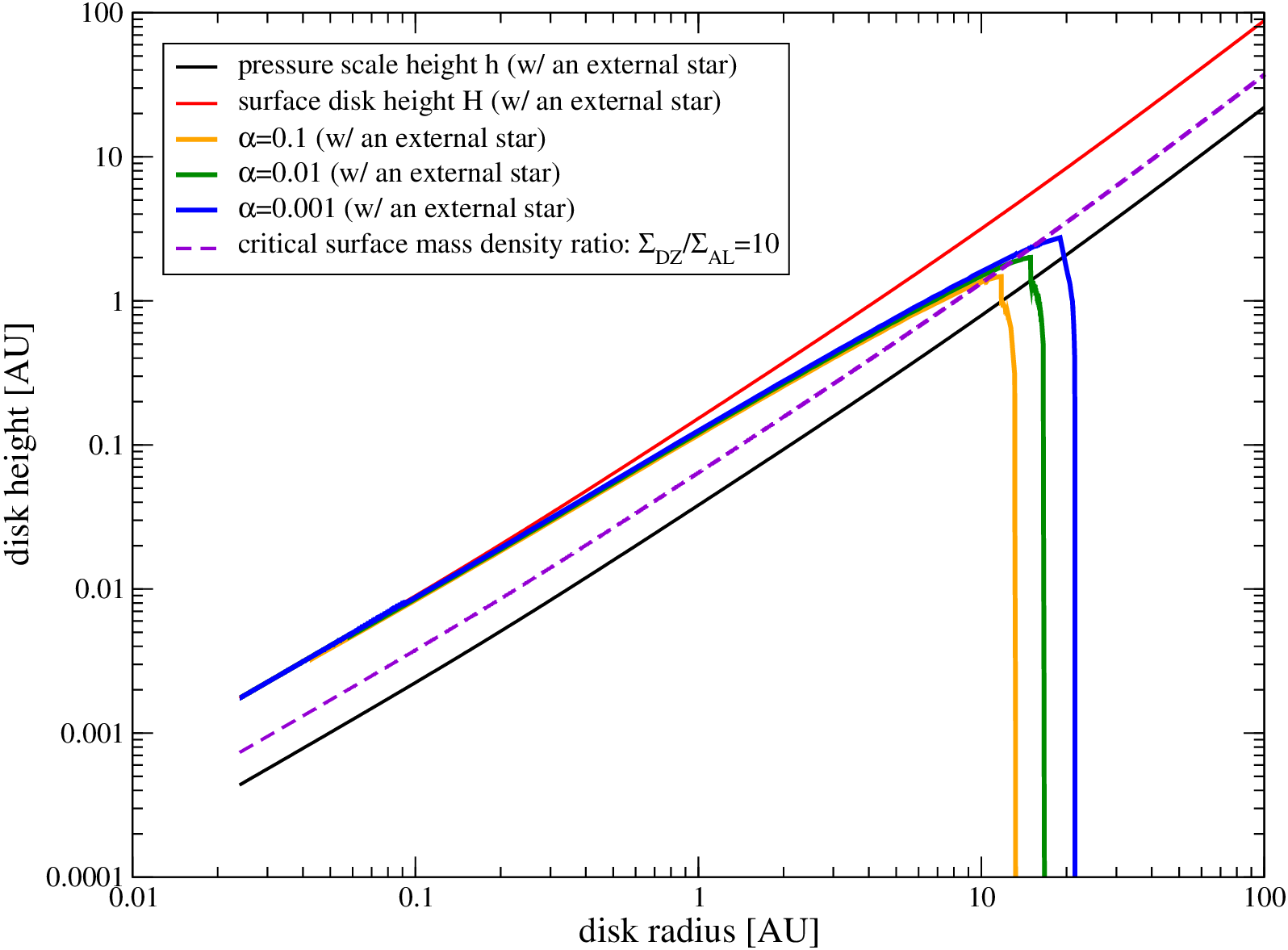}}
\end{picture}
\hspace{2cm}
\begin{picture}(5.5,6)
\scalebox{0.28}{
\includegraphics{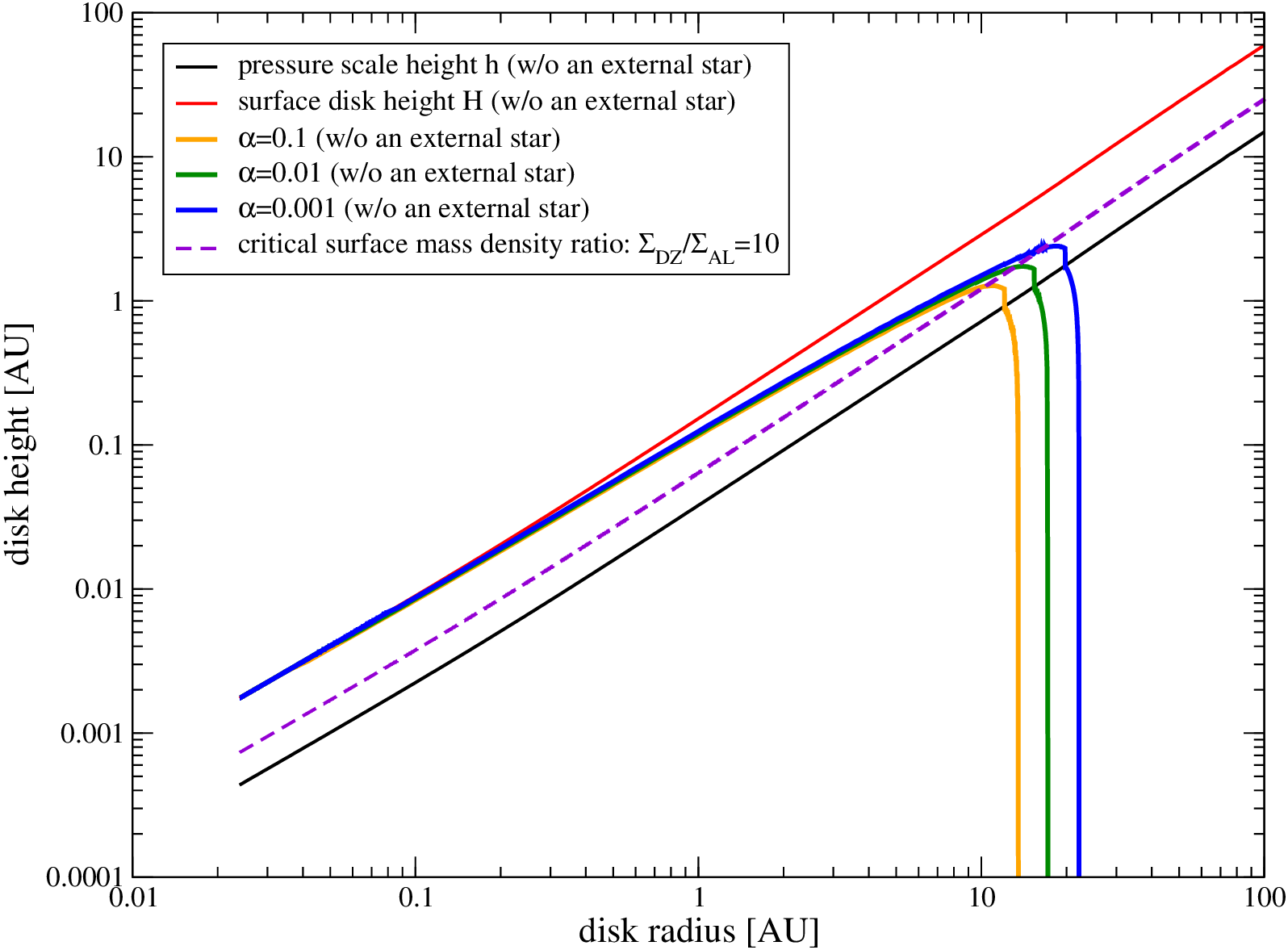}}
\end{picture}
\end{minipage} 
\hspace{1.5cm}
\parbox[t]{2cm}{{Total}}
\end{minipage}
\begin{minipage}[t]{18.5cm}
\hspace{0.5cm}
\begin{minipage}[h]{14cm}
\begin{picture}(5.5,6)
\scalebox{0.28}{
\includegraphics{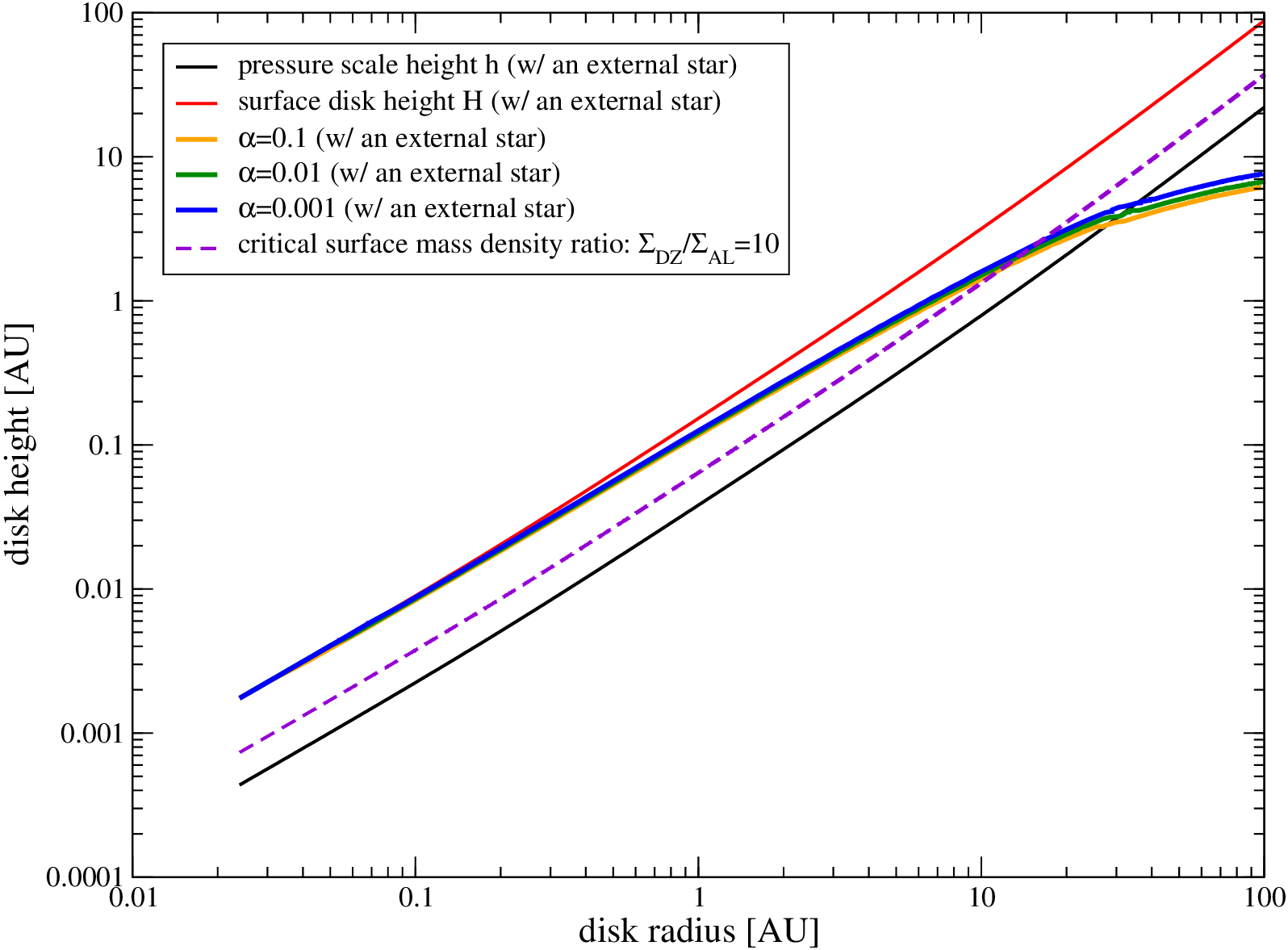}}
\end{picture}
\hspace{2cm}
\begin{picture}(5.5,6)
\scalebox{0.28}{
\includegraphics{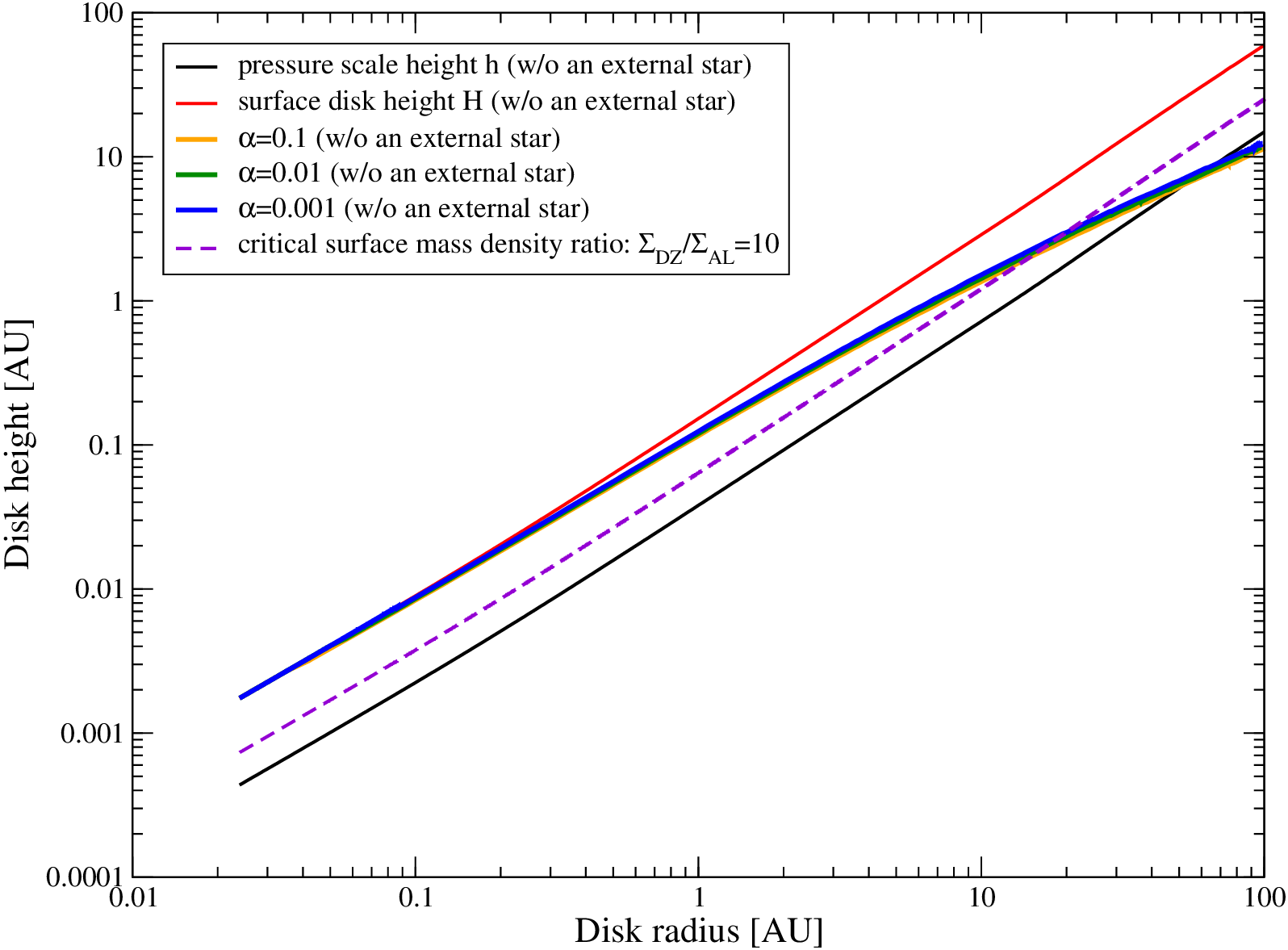}}
\end{picture}
\end{minipage}
\hspace{1.5cm}
\parbox[t]{2cm}{{X-rays}}
\end{minipage}
\begin{minipage}[t]{18cm}
\vspace{0.5cm}
\caption[dead2]{Comparing dead zones in clustered (left panels) and
isolated (right panels) environments.  For the upper panels, ionization
sources are X-rays from the magnetosphere of a central star and an
external star (for a disk in a clustered environment) as well as
cosmic rays, radioactive elements, and thermal ionization by alkali
ions.  For the lower panels, the ionization sources are X-rays from the
magnetosphere of a central star and an external star (for a disk in a
clustered environment).  Here, we choose \(\Sigma_{0}=10^3 \ {\rm g \
cm^{-2}}\) with \(Re_{M}=10^3\).  The uppermost line shows the surface
disk height; the lowermost line shows the pressure scale height, while
three curves indicate the dead zone boundaries for \(\alpha_{ss}=\)
0.1, 0.01, and 0.001 from inside to outside. The dashed line is where
\(\Sigma_{{\rm below}}/\Sigma_{{\rm above}}=10\). \label{fig2}}
\end{minipage}
\end{figure}

\begin{figure}
\unitlength1cm
\begin{minipage}[t]{18.5cm}
\hspace{3.5cm}
\parbox[t]{5.5cm}{{\(\Sigma_0=10^3 \ {\rm g \ cm^{-2}}\)}}
\hspace{2cm}
\parbox[t]{5.5cm}{{\(\Sigma_0=10^4 \ {\rm g \ cm^{-2}}\)}}
\end{minipage}
\begin{minipage}[t]{18.5cm}
\hspace{0.5cm}
\begin{minipage}[h]{14cm}
\mbox{} \\
\begin{picture}(5.5,6)
\scalebox{0.28}{
\includegraphics{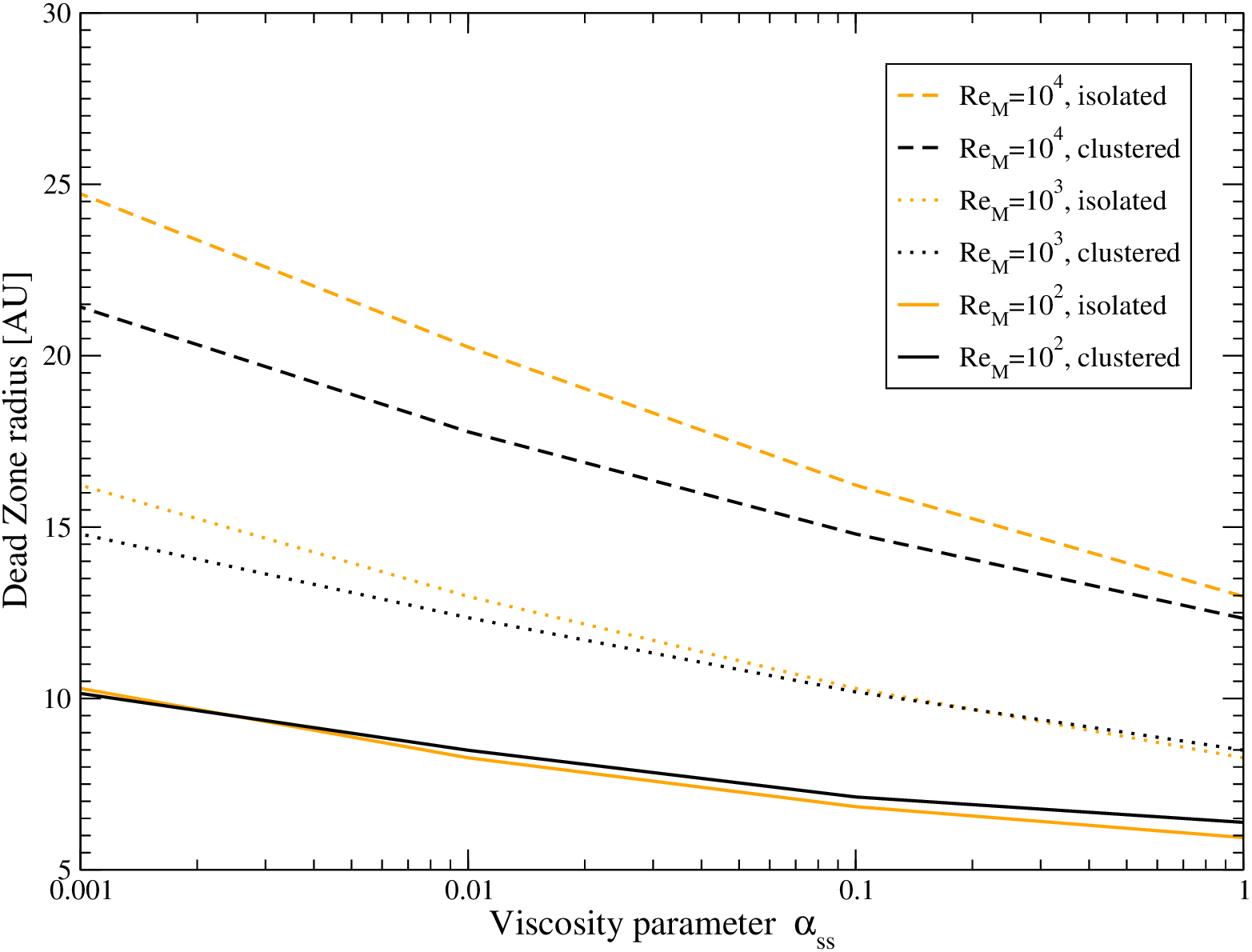}}
\end{picture}
\hspace{2cm}
\begin{picture}(5.5,6)
\scalebox{0.28}{
\includegraphics{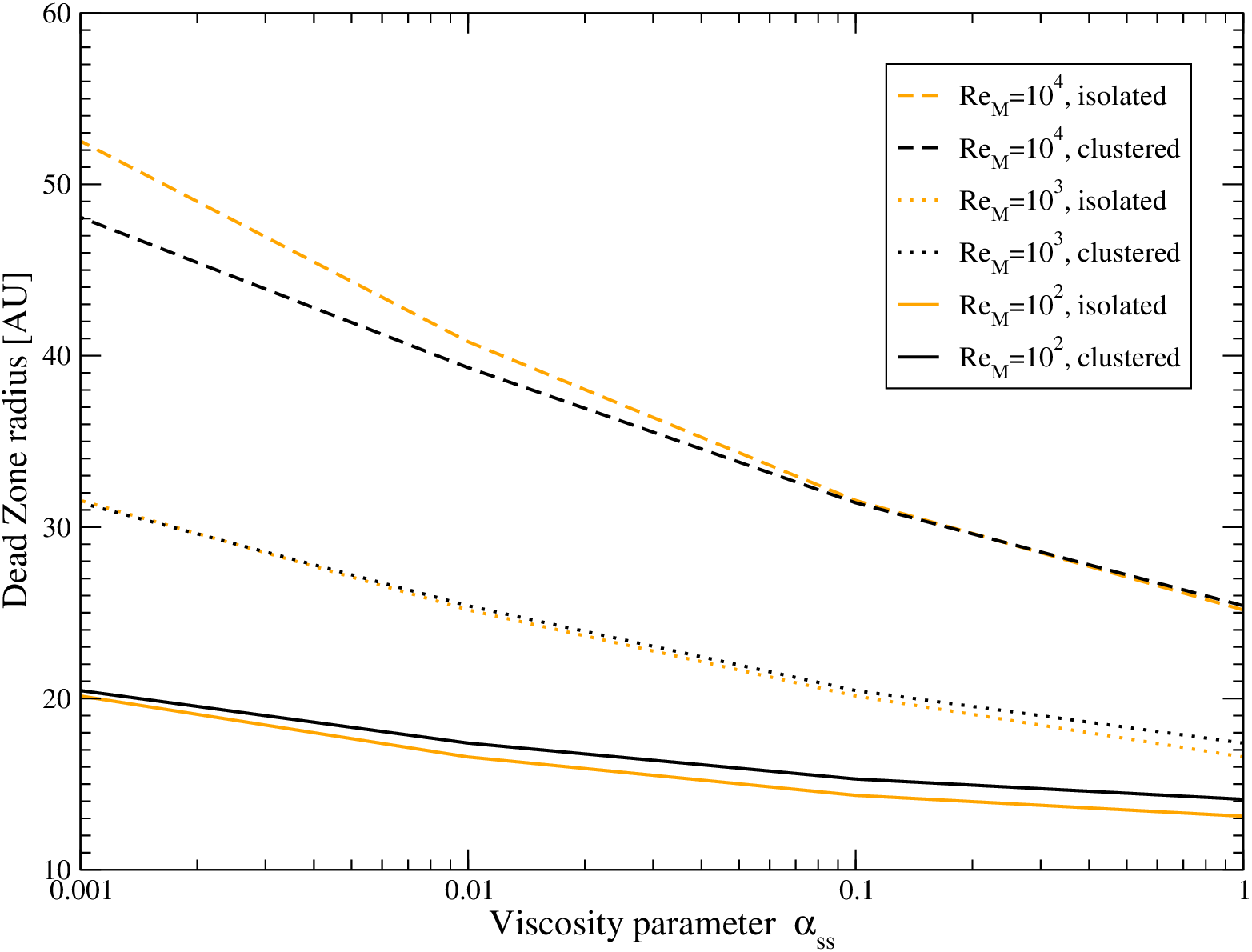}}
\end{picture}
\end{minipage}
\hspace{1.5cm}
\parbox[t]{2cm}{{Total}}
\end{minipage}
\begin{minipage}[t]{18.5cm}
\hspace{0.5cm}
\begin{minipage}[h]{14cm}
\begin{picture}(5.5,6)
\scalebox{0.28}{
\includegraphics{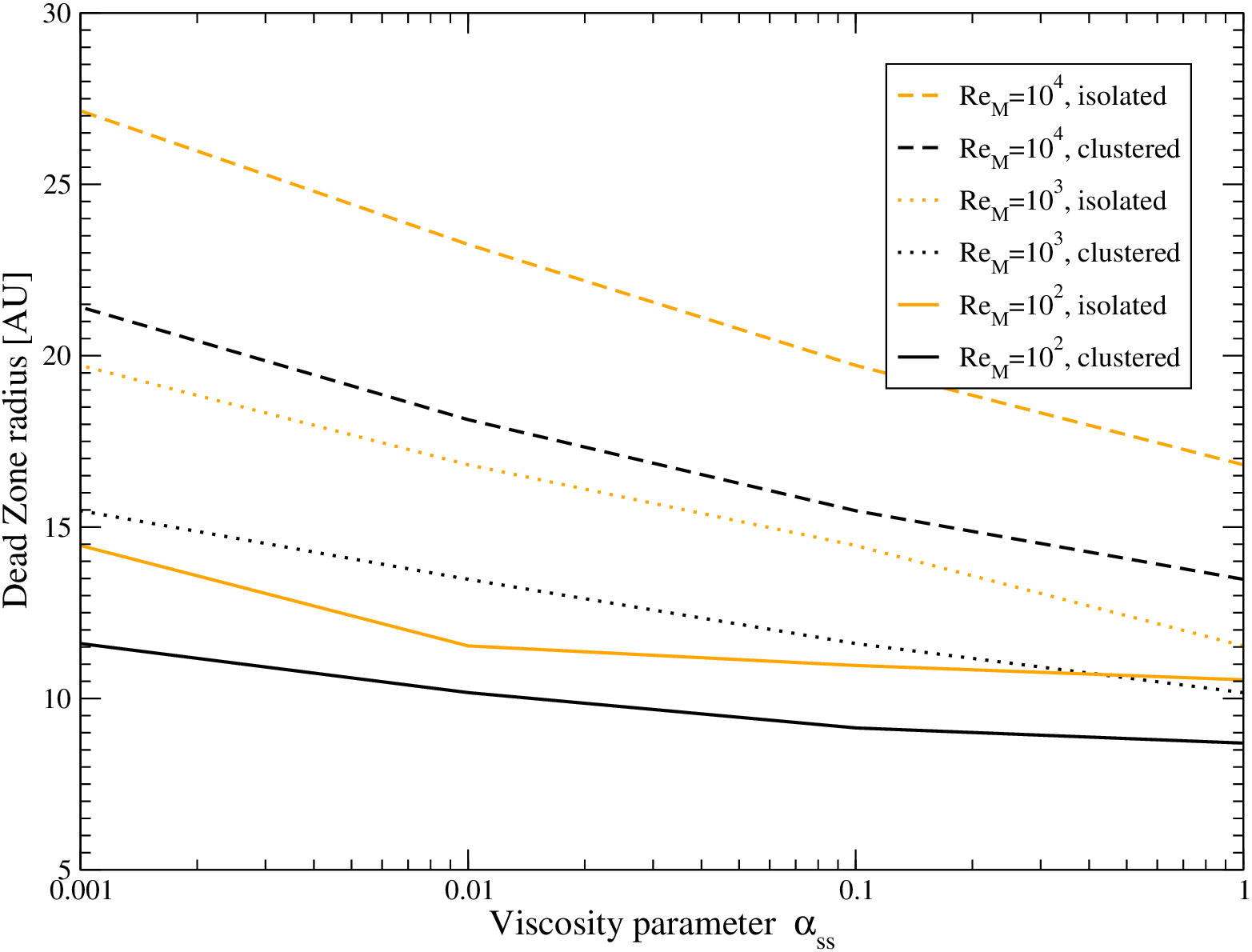}}
\end{picture}
\hspace{2cm}
\begin{picture}(5.5,6)
\scalebox{0.28}{
\includegraphics{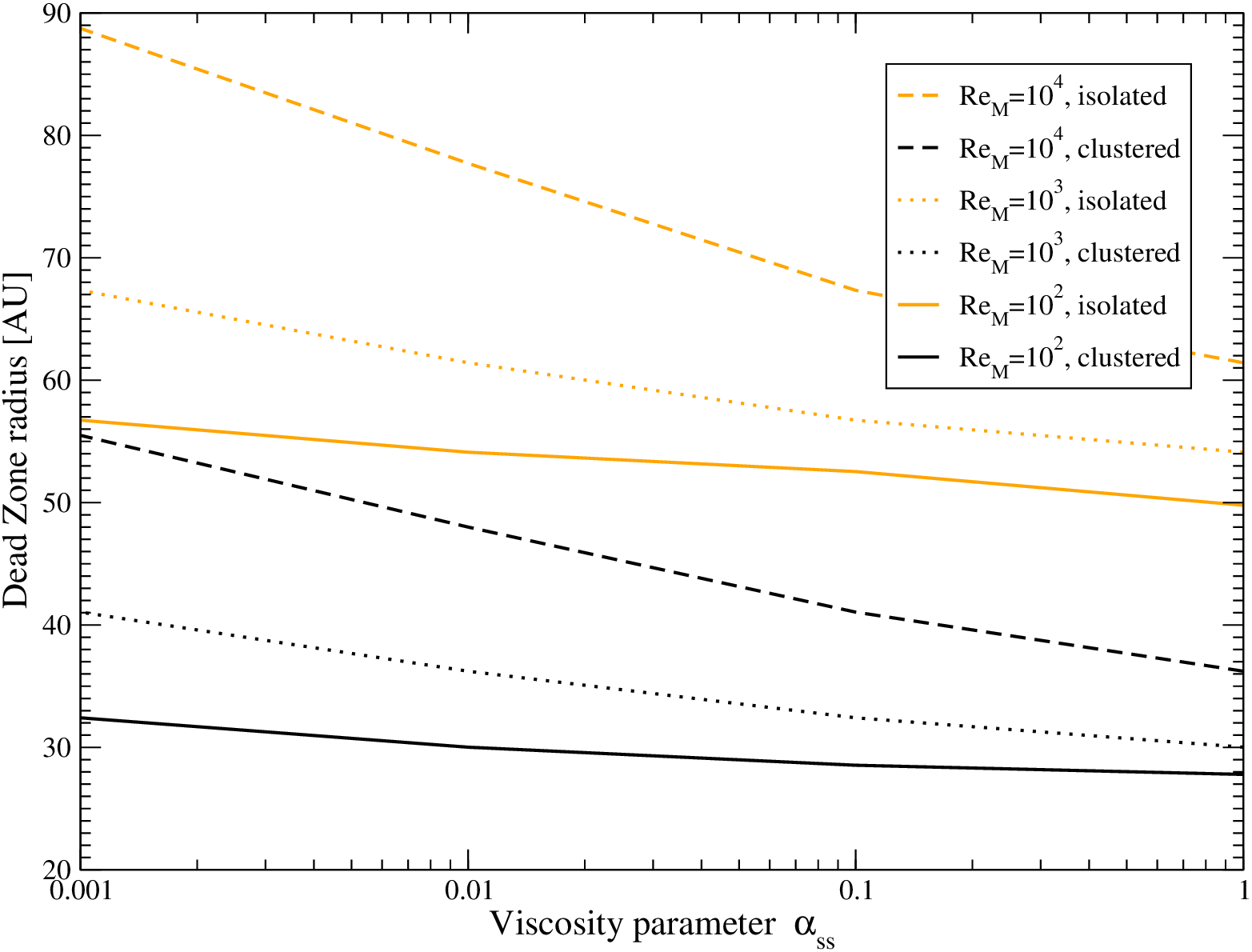}}
\end{picture}
\end{minipage}
\hspace{1.5cm}
\parbox[t]{2cm}{{X-rays}}
\begin{minipage}[t]{18cm}
\vspace{0.5cm}
\caption[rdead]{Comparing dead zone radii in clustered (heavy lines)
and isolated (light lines) environments.  Solid, dotted and dashed lines are for
\(Re_{M}=10^2, \ 10^3\), and \(10^4\) respectively.  The difference
between two environments are more apparent in the cases without cosmic
ray ionization (lower panels) compared to general cases (upper
panels).  Note the difference in scale of the y axes. \label{fig3}}
\end{minipage}
\end{minipage}
\end{figure}

Fig. \ref{fig2} shows dead zones in all representative cases of our
study -- disks in a cluster, isolated disks, and disks with dead zones
estimated by X-ray ionizations alone for each environment. 
Although we cover a large parameter space: the surface mass density of
\(\Sigma_0=10^3 - 10^4 \ {\rm g \ cm^{-2}}\), the magnetic Reynolds
number of \(Re_{M}=10^2 - 10^4\) and \(\alpha_{ss}=1.0-0.001\), we
only show the case of \((Re_{M}, \ \alpha_{ss})=(10^3, \ 0.1-0.001)\)
with \(\Sigma_0=10^3 \ {\rm g \ cm^{-2}}\) in this figure.
We will discuss the effects of different parameters below (see
Fig. \ref{fig3} as well).

Comparing disks in a cluster with isolated disks (upper two panels),
we find that dead zones stretch out to \(\sim 10-20\) AU in both
environments.
This is not so surprising because the effect of an external star and a
nebular environment is not dominant within \(\sim 10\) AU (see
Fig. \ref{fig1}).
The insensitivity of the dead zone size to thermal environments suggests
that protostellar disks in both environments have a similar transition
radius of disk's viscosities.

Comparing the dead zones estimated by the total ionization (upper
panels in Fig. \ref{fig2}) with those estimated by the X-ray
ionization (lower panels), we can see that cosmic rays have a large
effect on a dead zone shape at outer radii.
Disks ionized by both X-rays and cosmic rays (as well as other
sources) have fully magnetically active regions beyond \(\sim 10\) AU,
while disks ionized by X-rays alone are expected to be able to sustain
MRI only when there is a significant mass-mixing between active and
dead zones (note that the dashed line represents a critical surface
mass density ratio of the dead zone and the active layers below which
mass-mixing between these two zones is not negligible).
This is essentially due to the geometrical difference between these
two major ionizing sources, X-rays from a central star and cosmic
rays: X-rays are emitted from a stellar magnetosphere and therefore
penetrate disks with an angle of \(\ll 90^{\circ}\).
As a result, disks become more optically thick toward the outer part
of a disk for X-rays. 
Cosmic rays on the other hand, will propagate down to the disk
preferentially along disk's magnetic field lines that are orthogonal
to the disk's surface, hit the disk with an angle of \(\sim
90^{\circ}\) and experience a similar optical thickness throughout the
disk.

As described in MP05, we define edges of the dead zones as
intersections of a critical surface mass density ratio (dashed line)
with curved boundaries of dead zones. 
The dead zone edges obtained this way are plotted in Fig. \ref{fig3},
where we can compare parameters' effects on the dead zone sizes in
different environments and ionization sources.
In a standard parameter range \((Re_{M}, \ \alpha_{ss})=(10^3-10^4, \
0.1-0.001)\), dead zones estimated by the total ionization (upper two
panels) tend to be about \(<10-20\) \% larger for disks in isolated
environments. 
For \(Re_{M}\) and \(\alpha_{ss}\), the size of dead zones changes by
about a factor of 2 each over three orders of magnitude.
The most significant effect however comes from a surface mass density
of the disk at 1 AU \(\Sigma_0\), for which the dead zone's size
changes by the same factor, 2, over one order of magnitude. 

The difference in a dead zone size due to environments becomes more
apparent for disks ionized only by X-rays (lower two panels).
This is because of the extra X-ray ionization effect by a nearby
massive star in a clustered star-formation region. 
In a relatively heavy disk with \(\Sigma_0=10^4 \ {\rm g \ cm^{-2}}\),
dead zones of isolated disks are up to about a factor of 2 larger than
those of disks in a cluster.
For \(Re_{M}\) and \(\alpha_{ss}\), the size of dead zones changes by
about a factor of 2 and 1.5 respectively over three orders of
magnitude, while for \(\Sigma_0\), the dead zone size changes by a
factor of \(>3\) over just one order of magnitude.
Thus, the environmental effect becomes more important if cosmic rays
are excluded from the disks. 
Note however, that our estimated X-ray ionization rate for a clustered
environment is probably overestimated, since we have not taken account
of any extinction for X-rays traveling to the protostellar disk and
since the adopted X-ray luminosity for the massive external star is
likely to be an upper limit as mentioned earlier.

In all cases, the most important parameter for the dead zone size is
the surface mass density of the disk at 1 AU \(\Sigma_0\). 
Both \(Re_{M}\) and \(\alpha_{ss}\) also have a large effect on the
dead zone size, while the environmental effect on the dead zone seems
to be negligible unless the X-ray ionization from external sources is
very powerful.
\subsection{Clustered vs Isolated Star Formation -- Planetary
Gap-opening Masses}
\begin{figure}
\scalebox{0.3}{
\includegraphics{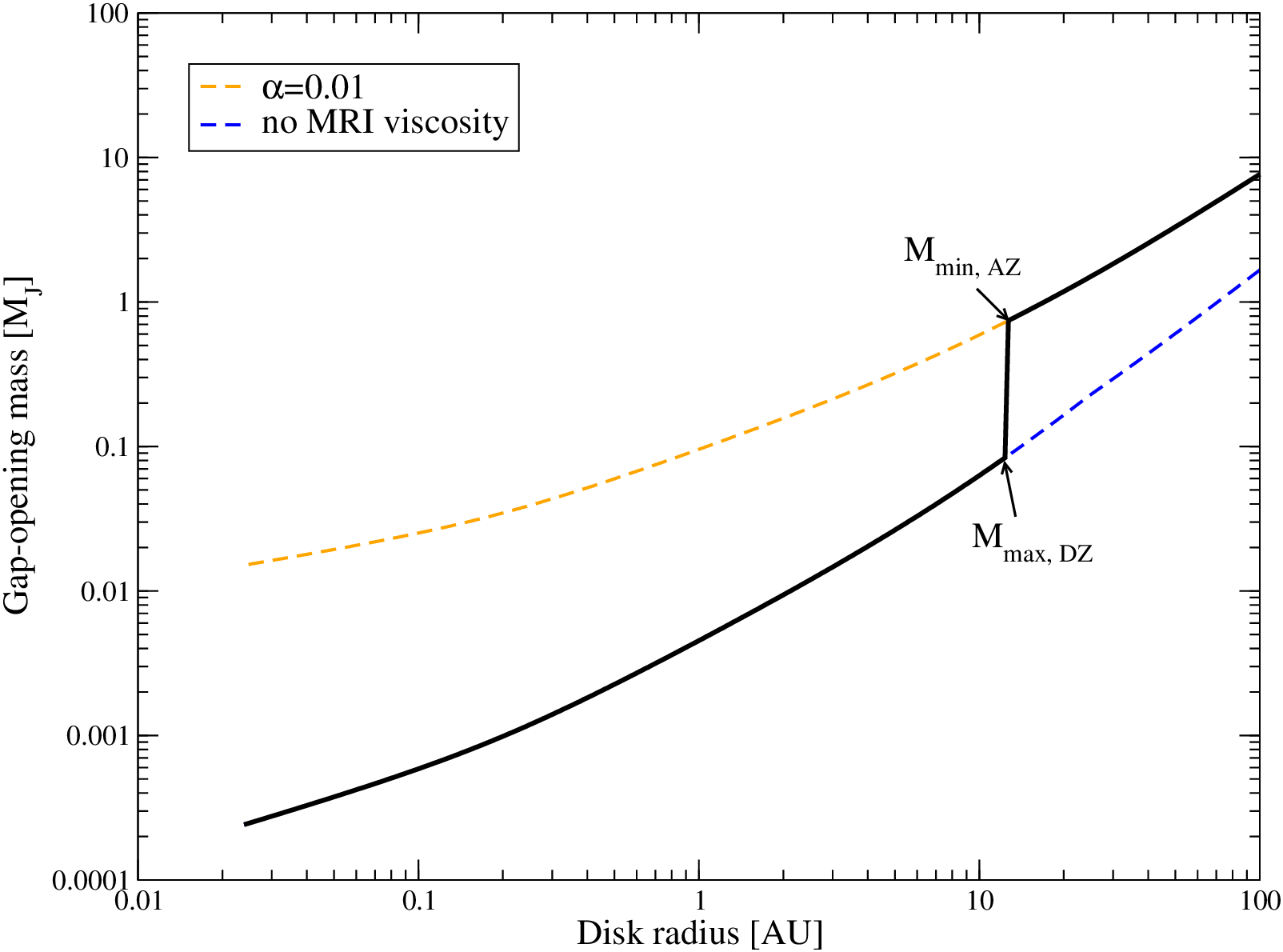}}
\caption[mass0]{Gap-opening masses for a disk exposed to an external
star which has the surface mass density, at 1 AU, of
\(\Sigma_{0}=10^3 \ {\rm g \ cm^{-2}}\).  The lower dashed line shows
the gap-opening mass for the case of no MRI viscosity, while the
upper dashed line shows the gap-opening masses for
\(\alpha_{ss}=0.01\).  For the magnetic Reynolds number
\(Re_{M}=10^3\), the fiducial minimum gap-opening mass throughout the
disk is shown in a heavy line. \label{fig4}}
\end{figure}

The general effect of a dead zone upon planetary masses is to create a
bimodal distribution.
Gap-opening masses will be sharply reduced as we move inwards through
the well-coupled outer region of a disk and encounter the dead zone.
This is shown in Fig. \ref{fig4} which plots the gap-opening mass as
a function of disk radius by assuming \(\alpha_{ss}=0.01\) in the
active region.
The figure shows the gap-opening mass for the case of \(\alpha_{ss}=0.01\)
throughout the disk (upper dashed curve), as well as that for the case of
inviscid disks (lower dashed curve).
The actual distribution is the heavy black curve, which follows the
low, inviscid curve in the dead zone, and then precipitously jumps up
to the upper curve in the active, well-coupled outer region of the disk.
The maximum mass of a planet in the disk's dead zone is \(M_{{\rm max,
DZ}}\sim 0.09 M_{J}\) in this example.  Just outside of the dead zone, the
minimum mass in the well-coupled active zone is  \(M_{{\rm min,
AZ}}\sim 0.79 M_{J}\). 

\begin{figure}
\unitlength1cm
\begin{minipage}[t]{19cm}
\hspace{3.5cm}
\parbox[t]{5.5cm}{{Disks in a cluster}}
\hspace{2cm}
\parbox[t]{5.5cm}{{Isolated Disks}}
\end{minipage}
\begin{minipage}[t]{19cm}
\hspace{0.5cm}
\begin{minipage}[h]{14cm}
\mbox{} \\
\begin{picture}(5.5,6)
\scalebox{0.28}{
\includegraphics{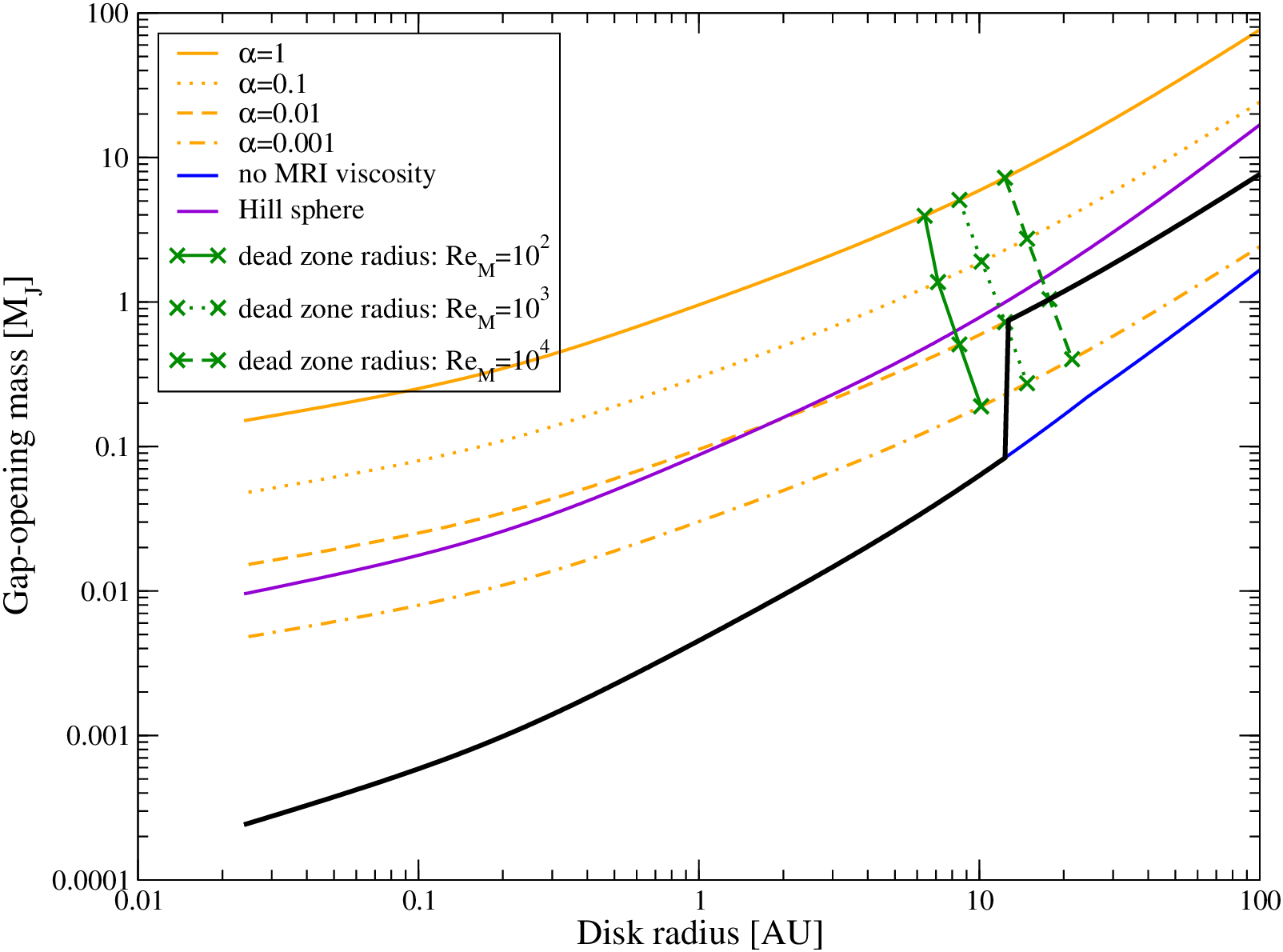}}
\end{picture}
\hspace{2cm}
\begin{picture}(5.5,6)
\scalebox{0.28}{
\includegraphics{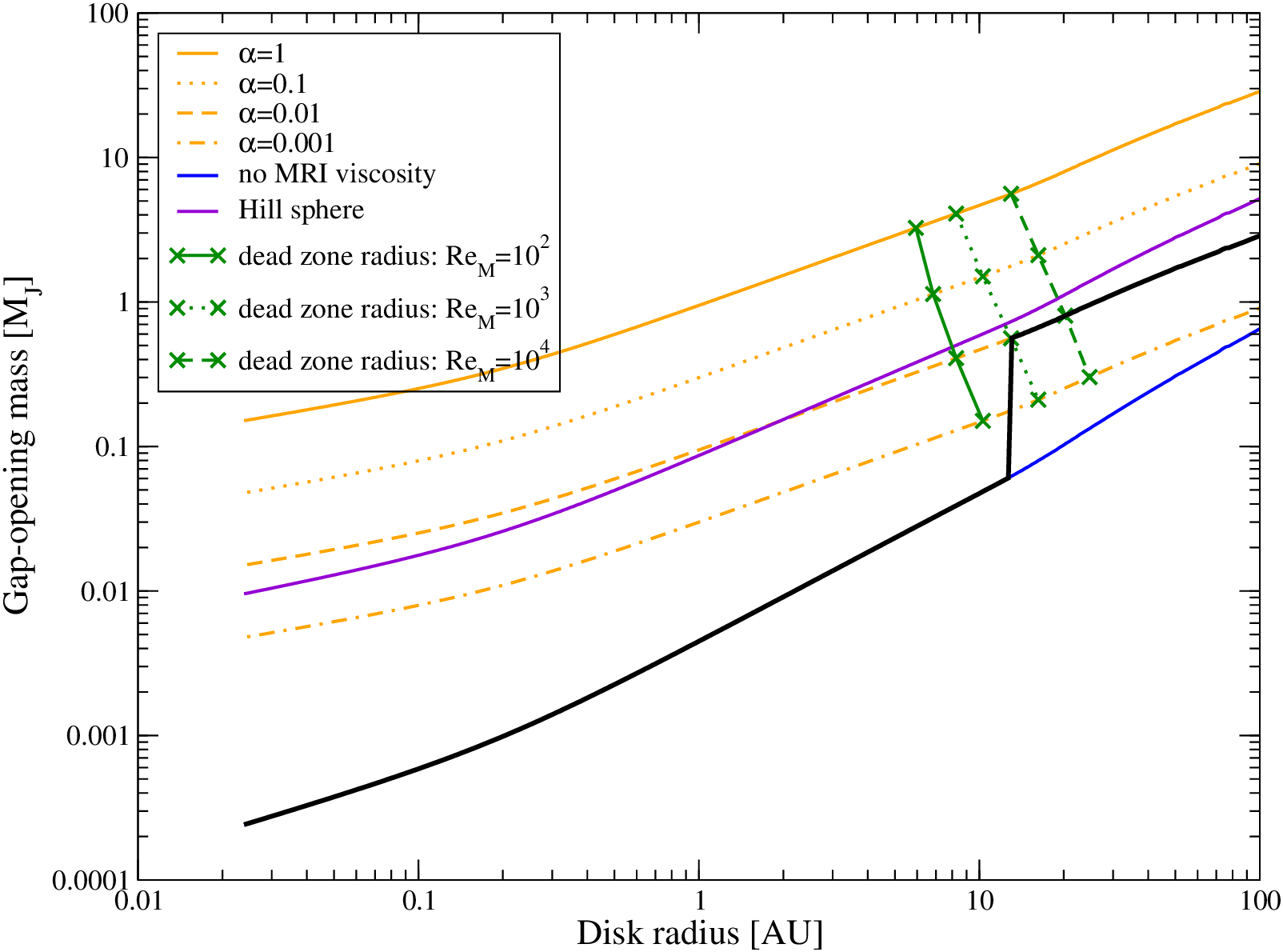}}
\end{picture}
\end{minipage} 
\hspace{1.5cm}
\parbox[t]{2.5cm}{{\(\Sigma_0=10^3 \ {\rm g \ cm^{-2}}\)}}
\end{minipage}
\begin{minipage}[t]{19cm}
\hspace{0.5cm}
\begin{minipage}[h]{14cm}
\begin{picture}(5.5,6)
\scalebox{0.28}{
\includegraphics{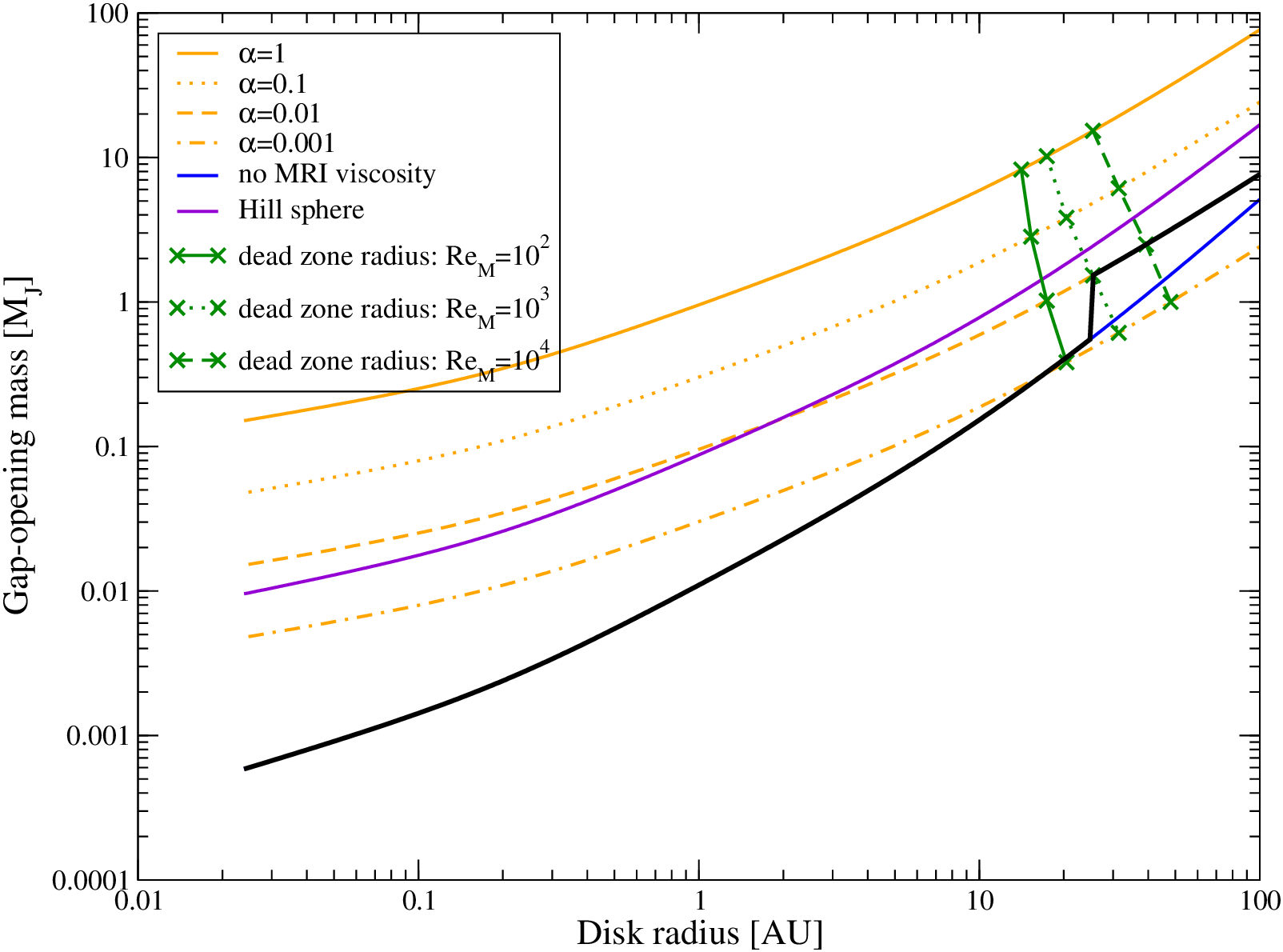}}
\end{picture}
\hspace{2cm}
\begin{picture}(5.5,6)
\scalebox{0.28}{
\includegraphics{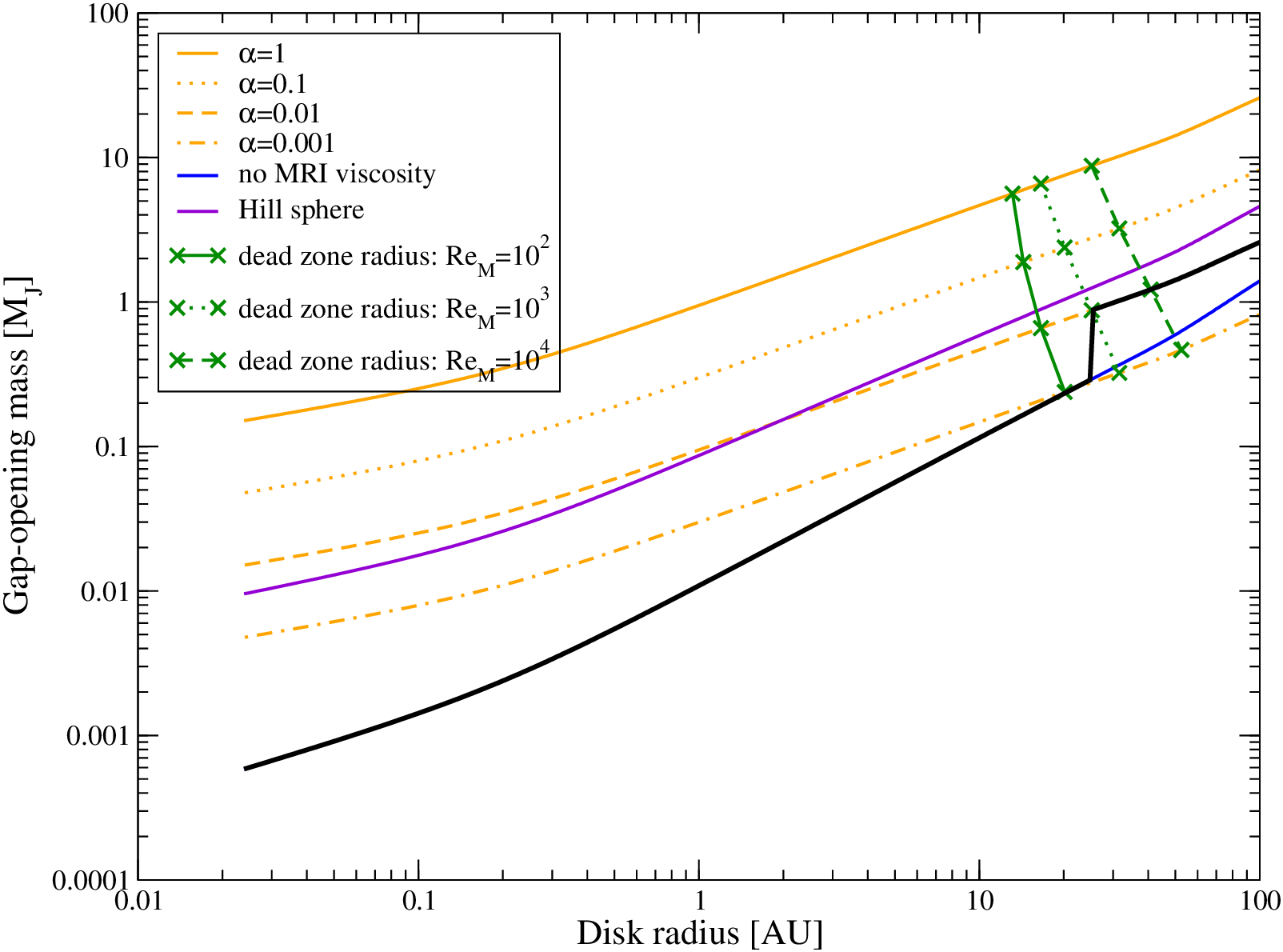}}
\end{picture}
\end{minipage}
\hspace{1.5cm}
\parbox[t]{2.5cm}{{\(\Sigma_0=10^4 \ {\rm g \ cm^{-2}}\)}}
\end{minipage}
\begin{minipage}[t]{18cm}
\vspace{0.5cm}
\caption[mass1]{Comparing dead zones in clustered (left panels) and
isolated (right panels) environments.  Dead zone radii are calculated
by using the total ionization rate in all cases.  For upper panels,
we assume the surface mass density at 1 AU of \(\Sigma_{0}=10^3 \ {\rm
g \ cm^{-2}}\), while for lower panels, \(\Sigma_{0}=10^4 \ {\rm g \
cm^{-2}}\). The lowermost line shows the gap-opening mass for the case
of no MRI viscosity, while the upper parallel lines show the
gap-opening masses for a different strength of magnetic field: the
solid line is for \(\alpha_{ss}=1\), the dotted line is for 0.1, the
dashed line is for 0.01, and the dot-dashed line is for 0.001.  The
dead zone for the magnetic Reynolds number, \(Re_{M}=10^2\), is inside
a solid line with crosses, that for \(10^3\) is inside a dotted line with
crosses, and that for \(10^4\) is inside a dashed line with crosses.  
The thick solid line traces our fiducial minimum gap-opening mass
throughout the disk for MRI ``viscosity,'' \(\alpha_{ss}=0.01\). \label{fig5}}
\end{minipage}
\end{figure}

We generalize this approach in Fig. \ref{fig5} which shows
gap-opening masses as a function of a disk radius for disks in a
cluster (left panels) and isolated disks (right panels).
In each panel, we plot the gap-opening mass lines for disks with
different, constant values of \(\alpha_{ss}\) (1.0, 0.1, 0.01 and
0.001) as well as Rafikov's inviscid case (see Equation (\ref{go2})).
The dead zone radius for a disk with each value of \(\alpha_{ss}\)
depends on the Reynolds number \(Re_{M}\) (MP05).
Thus, a cross appears on each \(\alpha_{ss}=\) constant curve, which
locates the dead zone radius for that particular disk model.
Since we show the data for these 3 different values of \(Re_{M}\), each
panel of this figure has 3 nearly vertical lines, which represent the
locus of dead zone radii for all disk models for each value of \(Re_{M}\).
These loci show that the dead zone radii are not highly sensitive to
the precise value of the magnetic Reynolds number \(Re_{M}\).

The reader may construct the predicted gap-opening masses for any disk
model (characterized by a value for \(\alpha_{ss}\) in the
well-coupled active zone, magnetic Reynolds number, surface mass
density, and ionization) by choosing the panels in Fig. \ref{fig5} to
construct a figure akin to that of Fig. \ref{fig4}.
One of the examples of gap-opening masses throughout the disk is
plotted in a heavy black line in each panel.
Also plotted is a mass inside a Hill radius (see Equation (\ref{go1}))
which is probably an upper limit for a gap-opening mass in an inviscid
disk.
This line also gives a rough estimate of planetary masses formed via disk
instability. 
If the disk's viscosity is larger than \(\sim 0.01\), these planets are
expected to keep on accreting gas after their formation.

An important implication of the results is that Jupiter or more
massive planets cannot be formed inside a dead zone (within \(\sim
10-20\) AU) through core accretion, because a gap opens for much lower
mass planets -- even for terrestrial planets.
These panels also indicate that an inviscid region in a typical
protostellar disk gives the same gap-opening masses as those expected
in a viscous disk with \(\alpha_{ss}=10^{-4} - 10^{-5}\). 
Therefore, disks with \(\alpha_{ss}=10^{-4} - 10^{-5}\) can be treated
as dead.

Another implication of Fig. \ref{fig5} is that super massive Jovian
planets are difficult to form.
For example, to obtain \(10 M_{J}\) within a reasonable radius \(\sim
40\) AU, the required viscosity parameter is \(\alpha_{ss}=0.1\) for a
clustered environment and \(\sim 1\) for an isolated environment. 
Therefore, we speculate that those massive planets may be formed
directly through the gravitational instability of the disk, or they
could accrete much more gas as they migrate.
Regarding the latter case, a moving planet is shown not to deplete its
feeding zone as much as a static planet \cite[e.g.][]{Rafikov02,Alibert04}.

\begin{figure}
\unitlength1cm
\begin{minipage}[t]{18.5cm}
\hspace{3.5cm}
\parbox[t]{5.5cm}{{\(\Sigma_0=10^3 \ {\rm g \ cm^{-2}}\)}}
\hspace{2cm}
\parbox[t]{5.5cm}{{\(\Sigma_0=10^4 \ {\rm g \ cm^{-2}}\)}}
\end{minipage}
\begin{minipage}[t]{18.5cm}
\hspace{0.5cm}
\begin{minipage}[h]{14cm}
\mbox{} \\
\begin{picture}(5.5,6)
\scalebox{0.28}{
\includegraphics{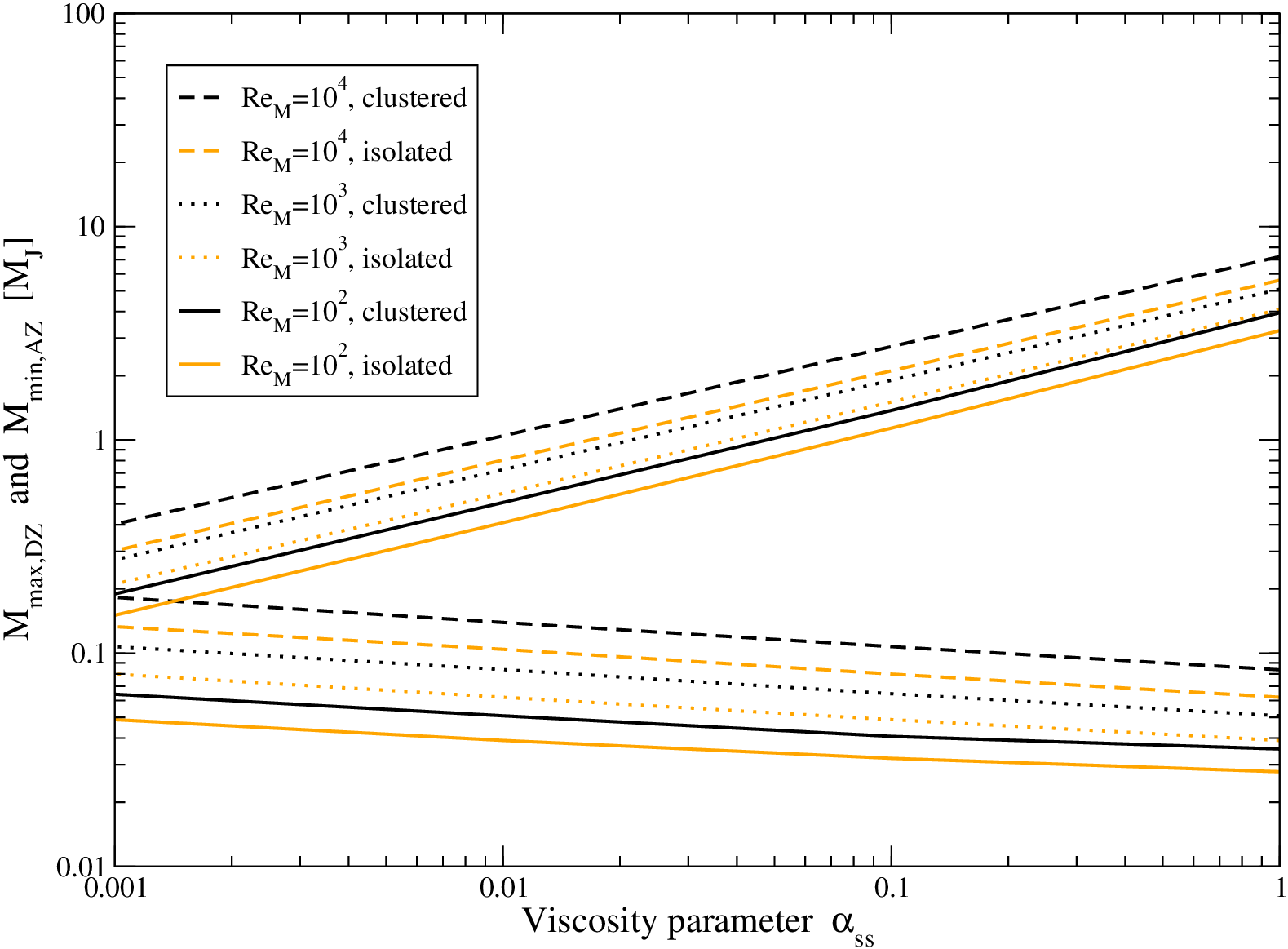}}
\end{picture}
\hspace{2cm}
\begin{picture}(5.5,6)
\scalebox{0.28}{
\includegraphics{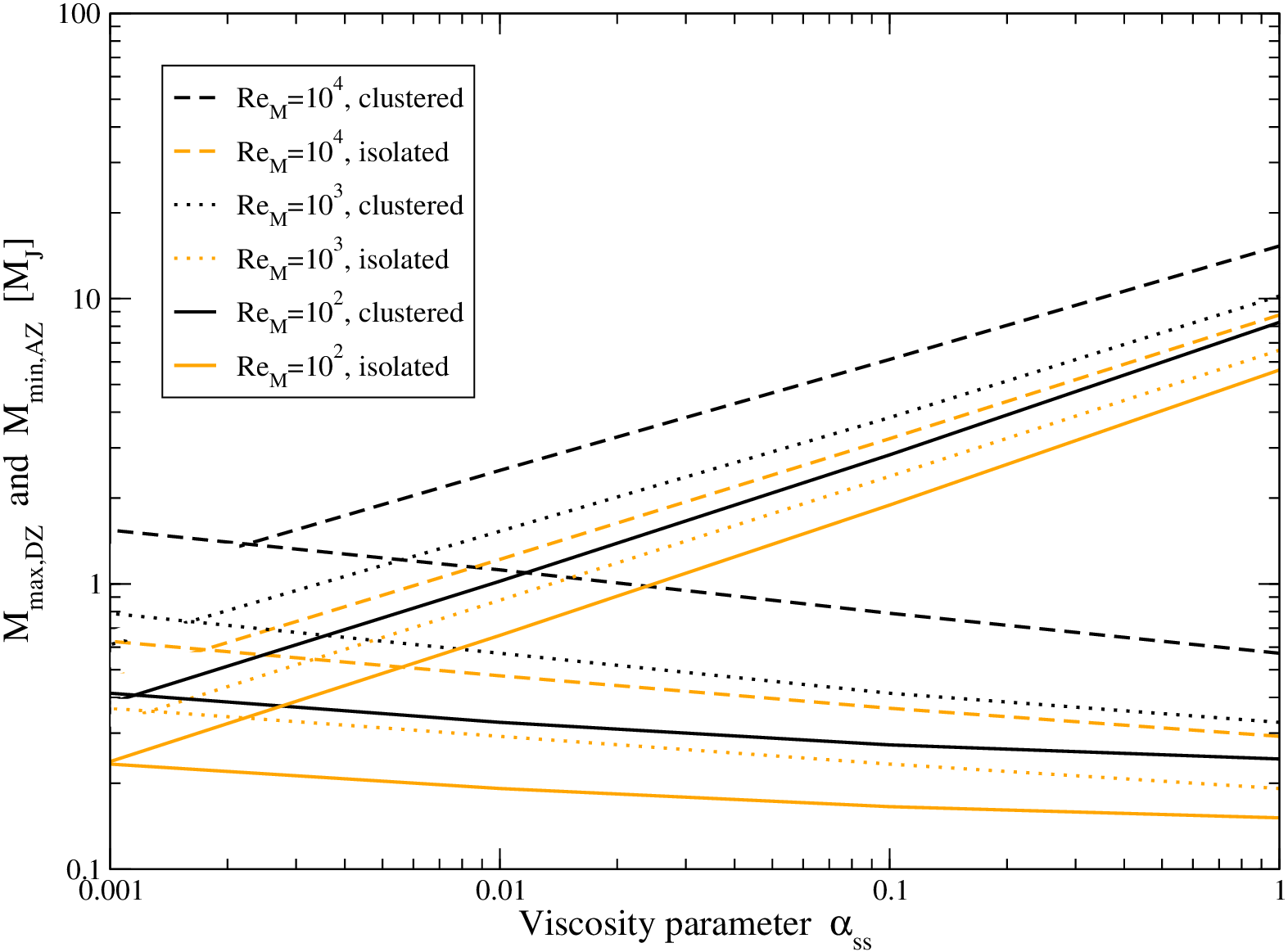}}
\end{picture}
\end{minipage}
\hspace{1.5cm}
\parbox[t]{2cm}{{Total}}
\end{minipage}
\begin{minipage}[t]{18.5cm}
\hspace{0.5cm}
\begin{minipage}[h]{14cm}
\begin{picture}(5.5,6)
\scalebox{0.28}{
\includegraphics{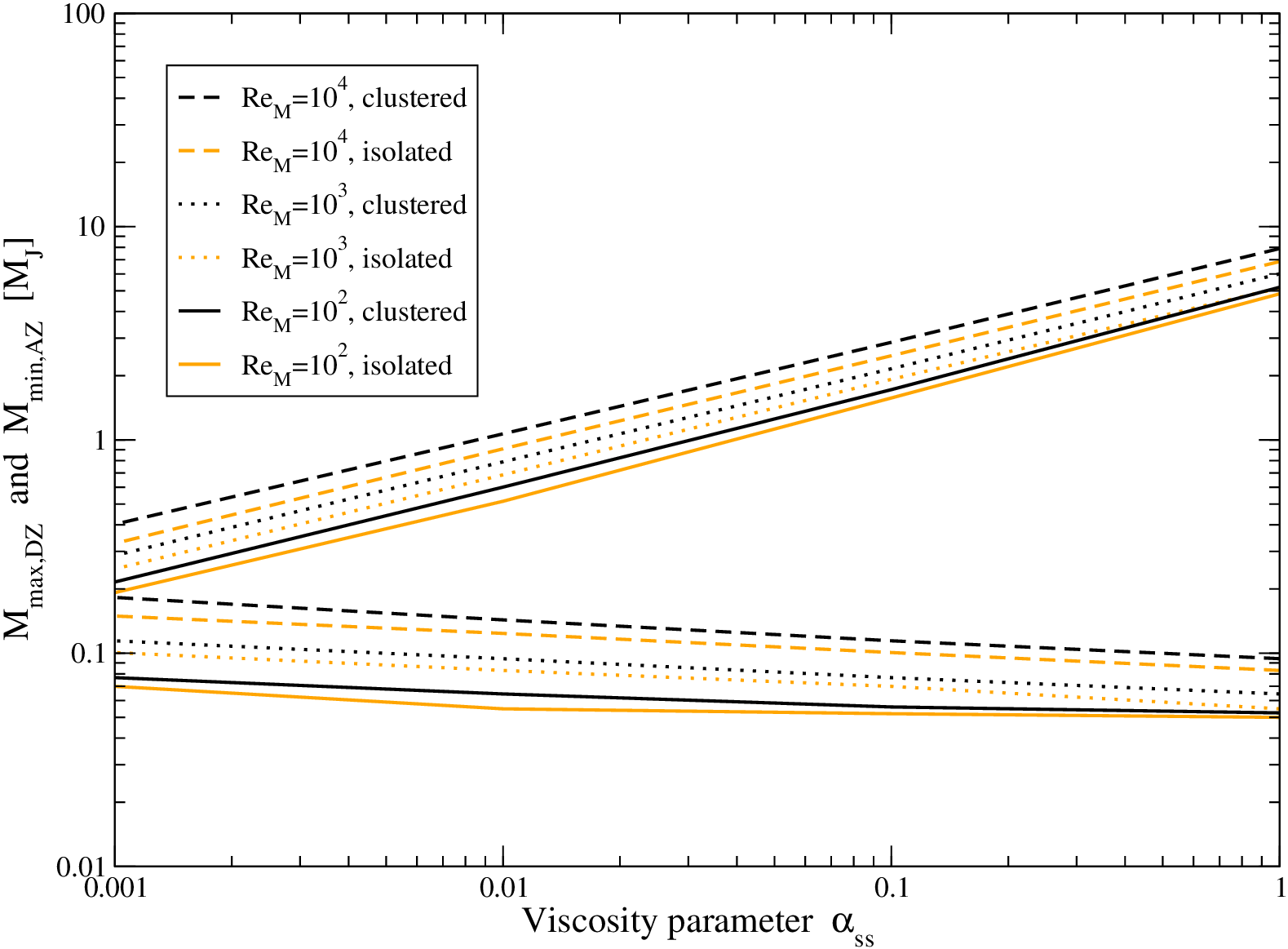}}
\end{picture}
\hspace{2cm}
\begin{picture}(5.5,6)
\scalebox{0.28}{
\includegraphics{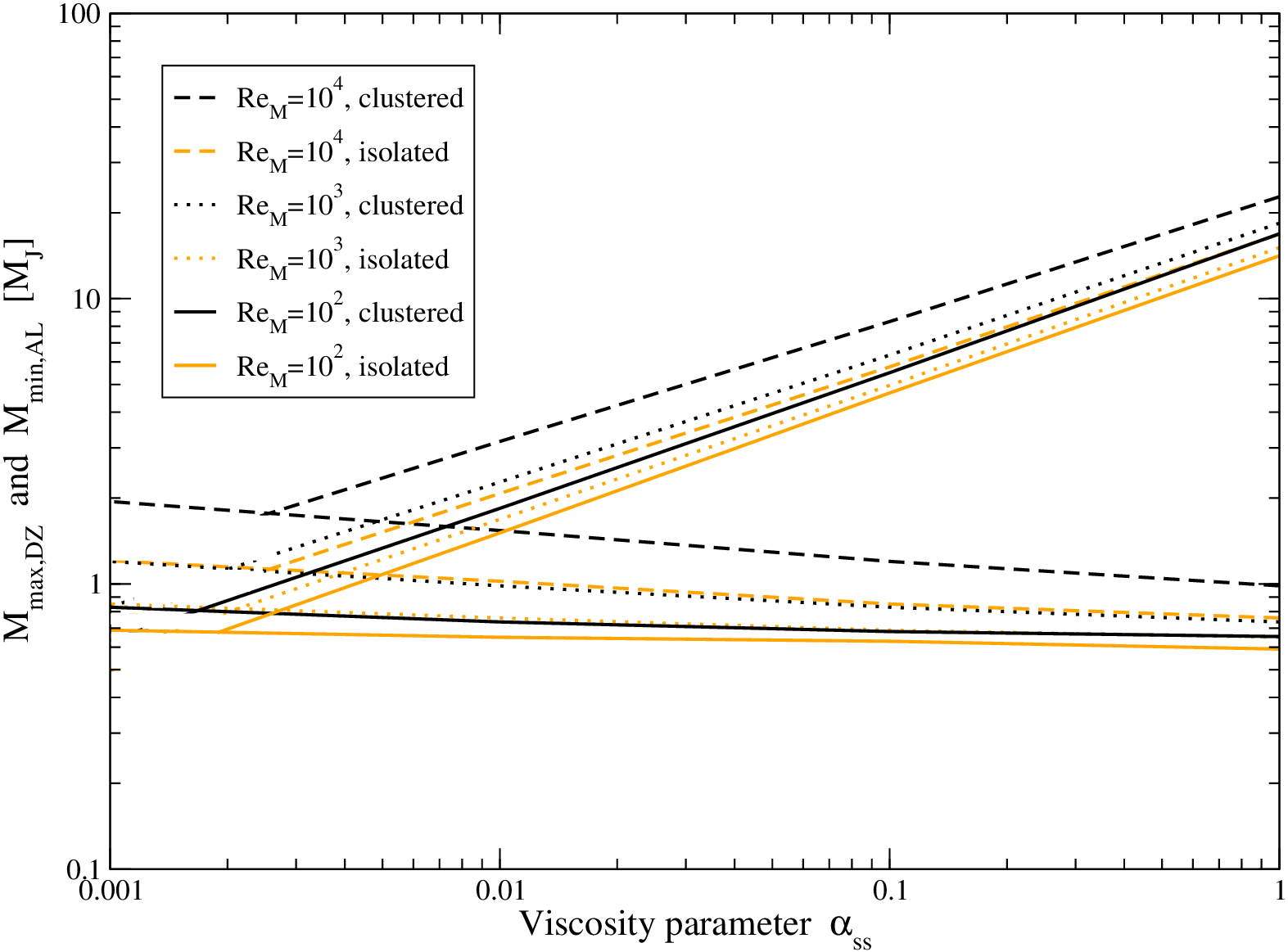}}
\end{picture}
\end{minipage}
\hspace{1.5cm}
\parbox[t]{2cm}{{X-rays}}
\begin{minipage}[t]{18cm}
\vspace{0.5cm}
\caption[mass_maxmin]{Comparing gap-opening masses just inside and
outside the dead zone radii in clustered (heavy lines) and isolated
(light lines) environments.  Solid, dotted and dashed lines are for
\(Re_{M}=10^2, \ 10^3\), and \(10^4\) respectively.  The difference
between two environments are more apparent in the cases without cosmic
ray ionization (lower panels) compared to general cases (upper
panels).  Note the difference in scale of the y axes. \label{fig6}}
\end{minipage}
\end{minipage}
\end{figure}

We saw in Fig. \ref{fig4} that the bimodal character of planetary masses
created by dead zones can be characterized by a large jump in
planetary masses; \(M_{{\rm max,DZ}} \rightarrow M_{{\rm min,AZ}}\).
Fig. \ref{fig6} shows this jump in gap-opening masses just inside and
outside the dead zone radii.
For each value of \(Re_{M}\) and \(\alpha_{ss}\), the dead zone radii seen in
Fig. \ref{fig3} can be combined with gap-opening masses in
Fig. \ref{fig5} to produce Fig. \ref{fig6}. 
These panels clearly show there is a jump in mass at the edge of a
dead zone.
Since dead zone sizes are more or less the same in both clustered and
isolated environments (see Fig. \ref{fig3}), and since gap-opening
masses around typical dead zone radius (a few tens of AU) are about
the same (see Fig. \ref{fig5}), these minimum and maximum gap-opening
masses at a dead zone radius are about the same for both clustered and
isolated star-forming environments.

For a heavier disk with \(\Sigma_0=10^4 \ {\rm g \ cm^{-2}}\) (right
panels), the ``minimum'' and ``maximum'' gap-opening mass lines
intersect for a small viscosity parameter of \(\alpha_{ss} \sim 10^{-3}\).
This is because gap-opening masses for an inviscid disk exceeds those
for a viscous disk for a sufficiently small value of \(\alpha_{ss}\)
(see the intersection of the lower two curves in the lower panels of
Fig. \ref{fig5}).
Therefore, in a moderately heavy disk with \(\alpha_{ss} \leq
10^{-3}\), there is no jump in gap-opening masses. 

Environmental effects on gap-opening masses can also be seen in these
heavier mass disk cases.
In the upper right panel of Fig. \ref{fig6}, where the dead
zones are calculated by the total ionization, gap-opening masses are
significantly different compared to the corresponding case for
\(\Sigma_0=10^3 \ {\rm g \ cm^{-2}}\) (upper left panel), despite the
fact that the dead zone sizes are about the same in both environments
in both cases.
This implies the difference in gap-opening masses between two
environments seem more significant in a heavier disk. 
This is because dead zone radii for \(\Sigma_0=10^4 \ {\rm g \
cm^{-2}}\) is about \(20-50\) AU, while those for \(\Sigma_0=10^3 \ {\rm g \
cm^{-2}}\) is about \(10-25\) AU, and because environmental effects
kick in only beyond \(\sim 10\) AU.  
Another example is a heavy disk ionized only by X-rays (lower right
panel).
Here, gap-opening masses in two environments give more or less the same values.
However, note that we are comparing these masses at very different dead
zone radii -- for example, for \((Re_{M}, \ \alpha_{ss})=(10^3, \
0.01)\), the dead zone radius is 36 AU for a clustered environment and
61 AU for an isolated environment.  
Therefore, these results imply that gap-opening masses in an isolated
environment is changing less sharply, as we can see in Fig. \ref{fig5}.

\begin{figure}
\unitlength1cm
\begin{minipage}[t]{18.5cm}
\hspace{0.5cm}
\begin{minipage}[h]{14cm}
\begin{picture}(5.5,6)
\scalebox{0.28}{
\includegraphics{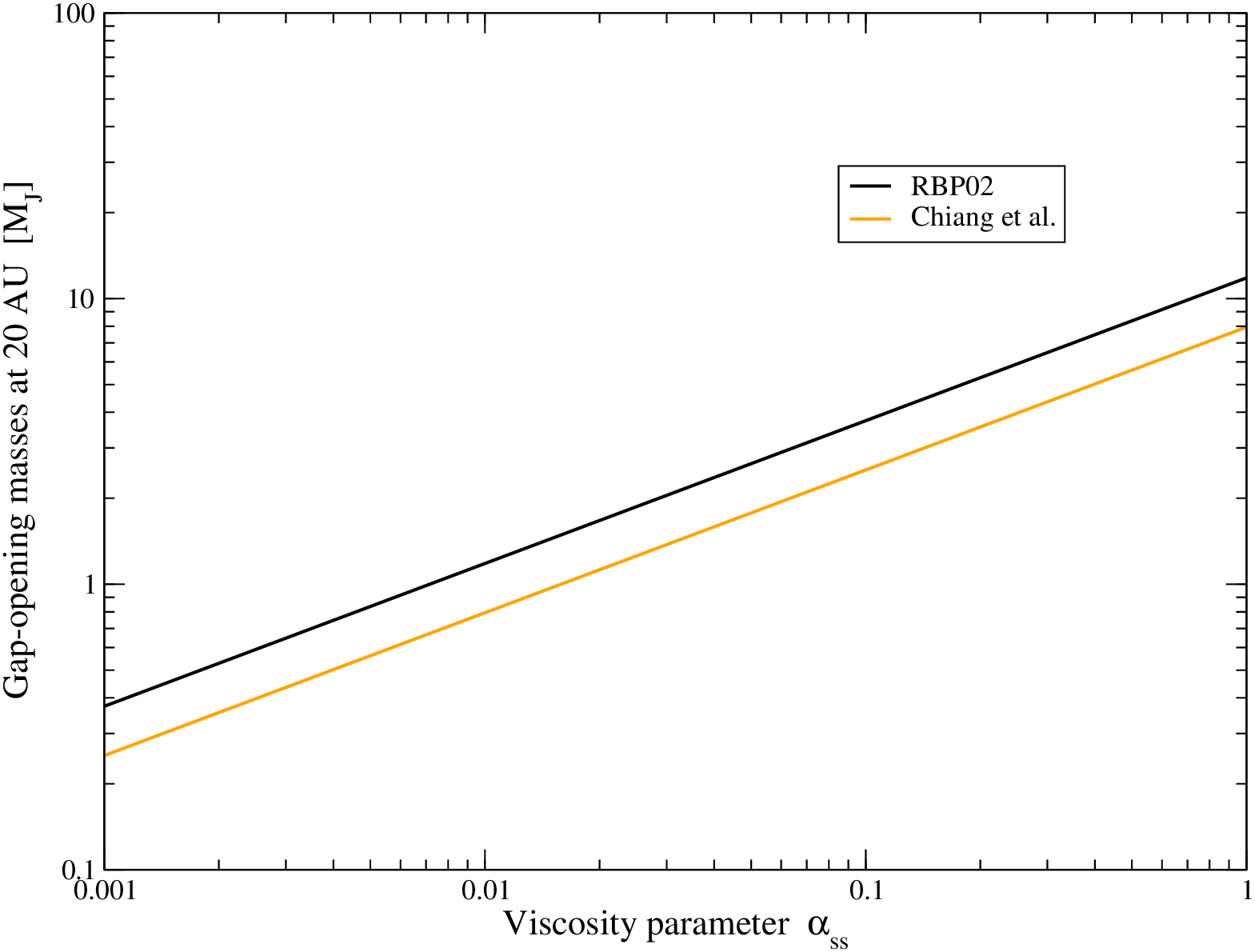}}
\end{picture}
\hspace{2cm}
\begin{picture}(5.5,6)
\scalebox{0.28}{
\includegraphics{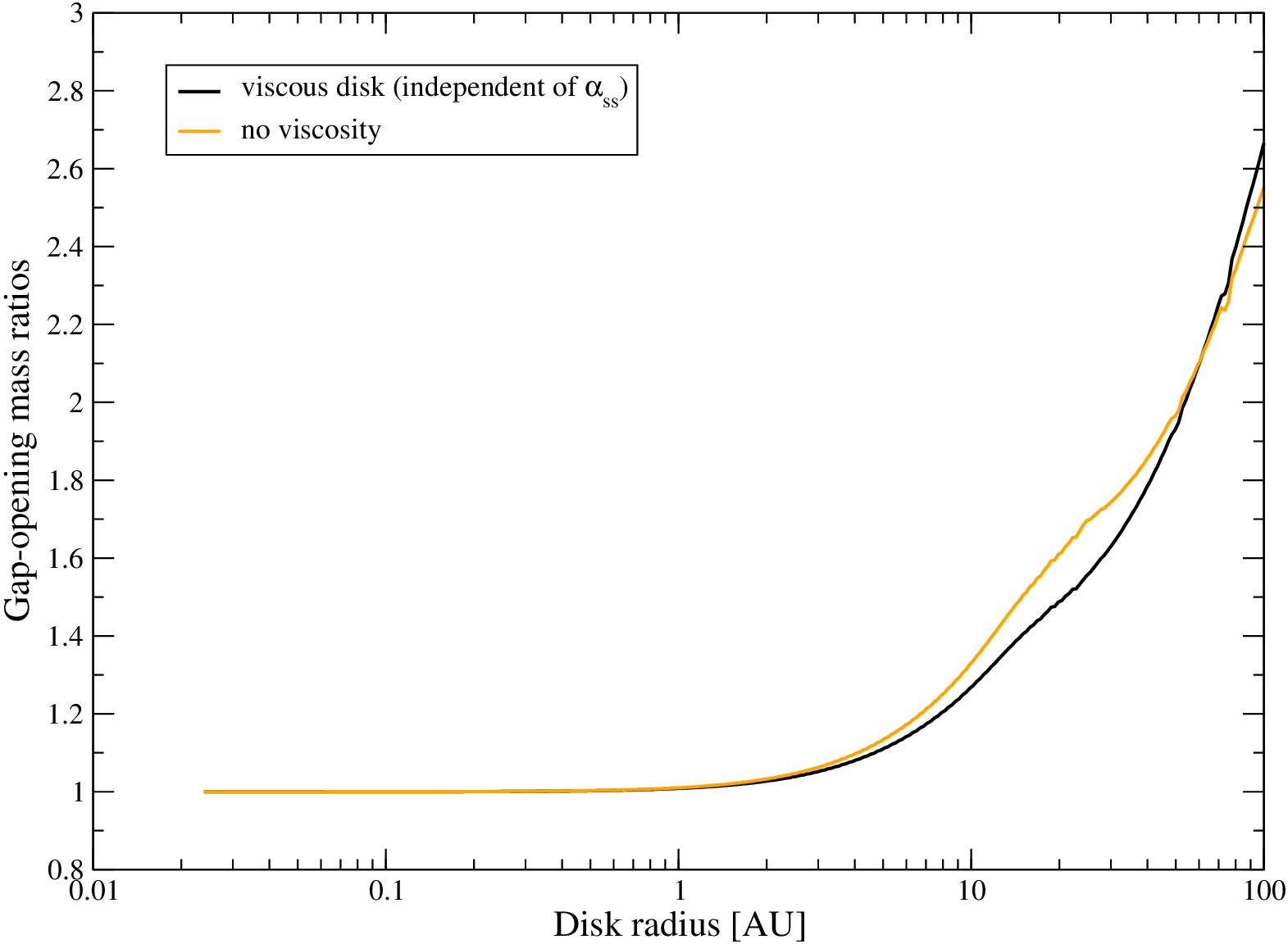}}
\end{picture}
\end{minipage}
\end{minipage}
\begin{minipage}[t]{18cm}
\vspace{0.5cm}
\caption[mass20au]{Left: comparison of gap-opening masses at 20 AU in
clustered (heavy line) and isolated (light line) environments.  Note
that the magnetic Reynolds number does not affect the gap-opening
masses.  Right: gap-opening mass ratios of isolated and clustered
star-forming environments.  Heavy line is for a viscous disk (see
Equation (\ref{go3})) and light line is for an inviscid disk (see
Equation (\ref{go2})).  Note that the results for a viscous disk are
independent of a viscosity parameter \(\alpha_{ss}\). \label{fig7}}
\end{minipage}
\end{figure}

In the left panel of Fig. \ref{fig7}, we compare gap-opening
masses at 20 AU in clustered and isolated star-forming regions.
The gap-opening masses in a clustered environment are about a factor of
1.5 larger than the ones in an isolated environment.
For standard viscosity parameters \(\alpha_{ss}=10^{-3} - 0.1\), the
gap-opening masses change from roughly a Saturn mass to a few Jupiter
masses at 20 AU. 
This indicates the importance of the viscosity parameter on a
gap-opening mass.

In the right panel of Fig. \ref{fig7}, we plot the gap-opening
mass ratios of clustered and isolated environments for both viscous
(heavy line, see Equation (\ref{go3})) and inviscid (light line, see
Equation (\ref{go2})) cases.  
Note that the result for a viscous case is independent of an actual
value of a viscosity parameter.
Different star-forming regions change a planetary gap-opening mass by
a factor of \(\sim 1.2 - 2.8\) beyond 10 AU in both viscous and
inviscid disks. 
This result further confirms that the environmental difference is
probably not a large factor in determining gap-opening masses. 
\subsection{Grain Evaporation in the Inner Disk}
\begin{figure}
\scalebox{0.3}{
\includegraphics{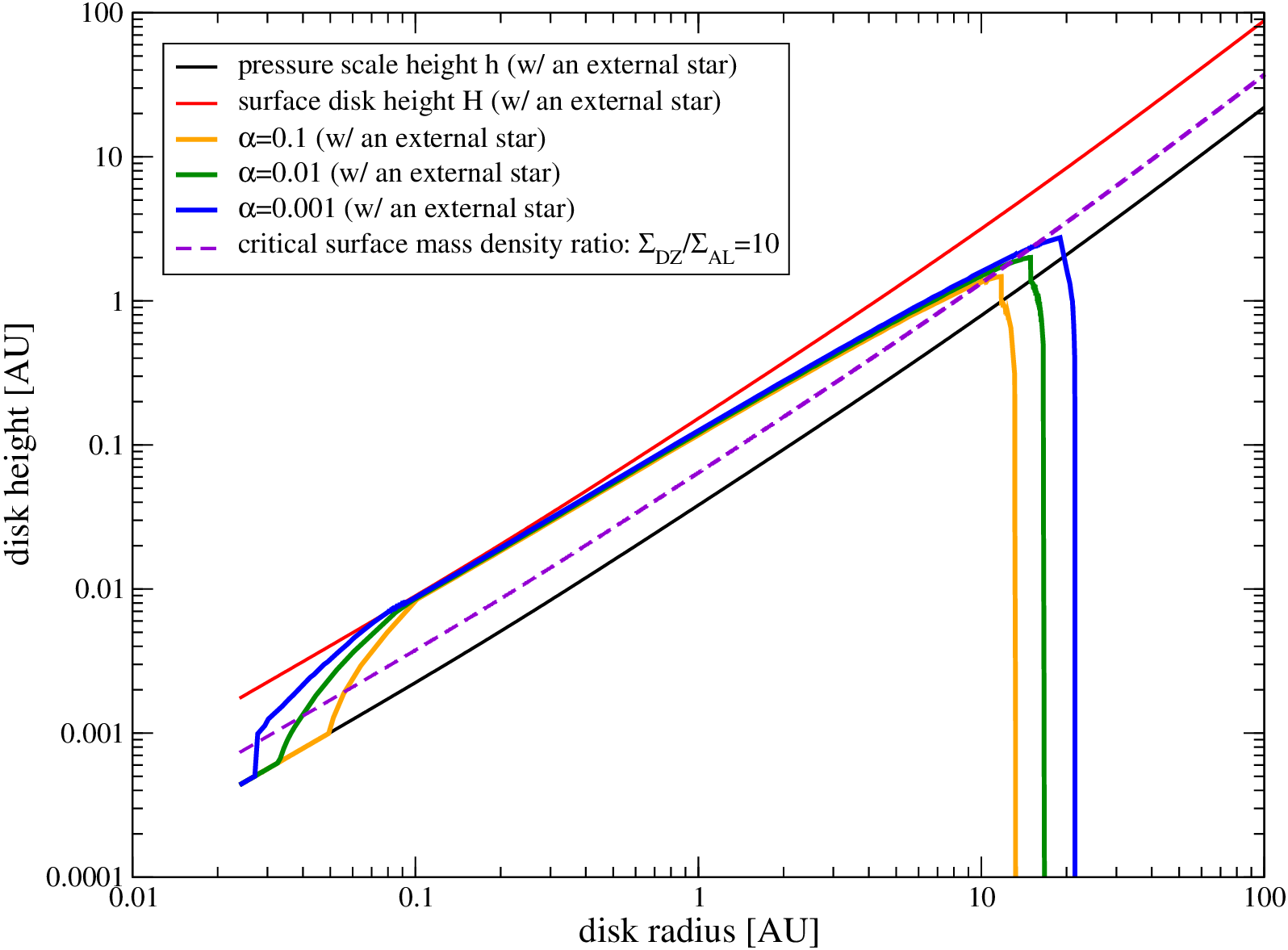}}
\caption[dead2]{The effect of grain evaporation on the sizes of dead zones
calculated for a protostellar disk exposed to an external star.  The
ionization sources are X-rays from the magnetosphere
of a central star and an external star as well as cosmic rays, radioactive
elements, and thermal ionization by alkali ions.  We use
\(\Sigma_{0}=10^3 \ {\rm g \ cm^{-2}}\) and \(Re_{M}=10^3\) with
\(\alpha_{ss}=\) 0.1, 0.01, and 0.001.  For the explanation of each
line, see Fig. \ref{fig2}. \label{fig8}}
\end{figure}

\begin{figure}
\scalebox{0.3}{
\includegraphics{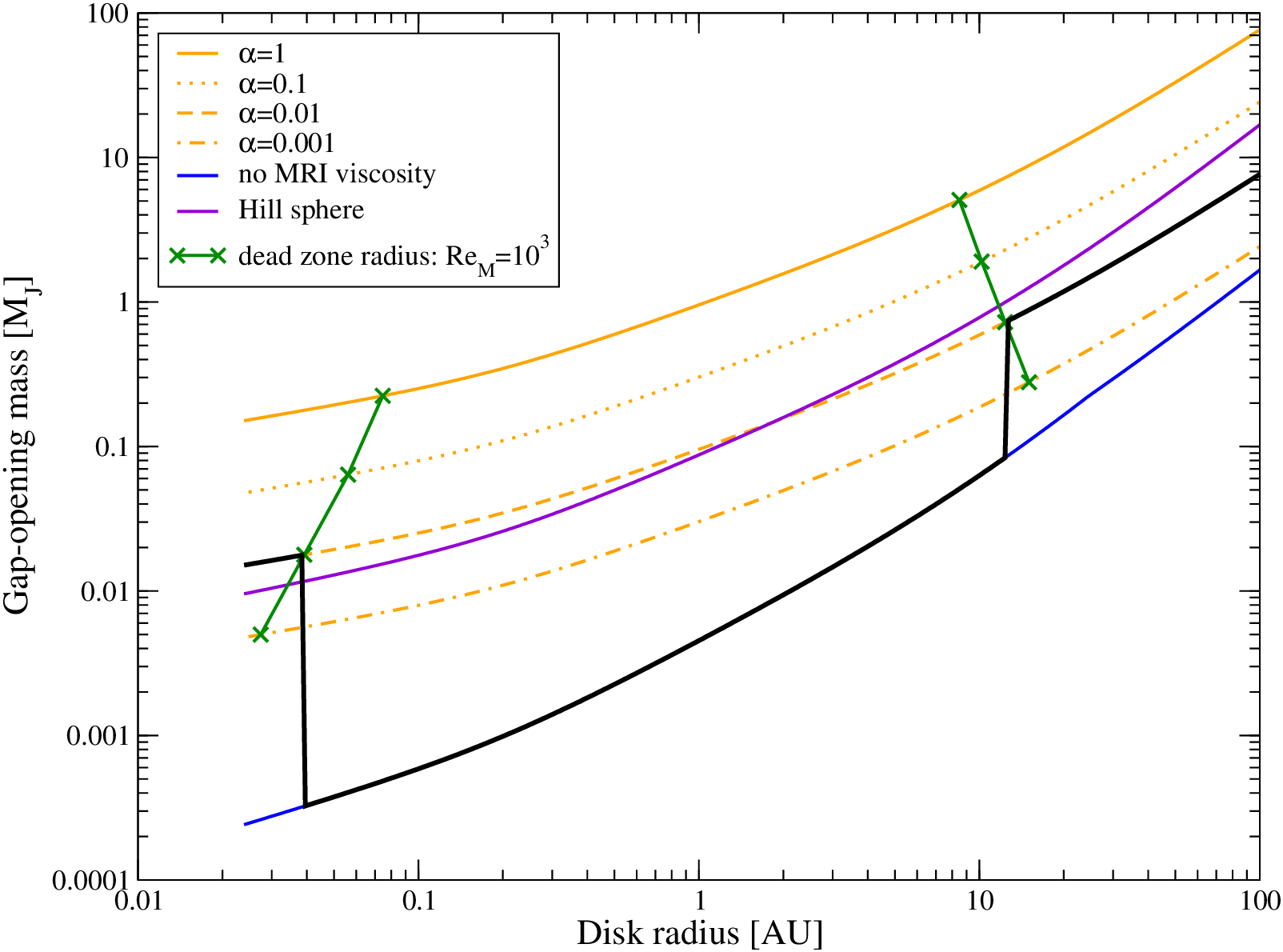}}
\caption[mass2]{The effect of grain evaporation on gap-opening masses
for a disk exposed to an external star which has the surface mass
density, at 1 AU, of \(\Sigma_{0}=10^3 \ {\rm g \ cm^{-2}}\).  For the
explanation of each line, see Fig. \ref{fig5}. \label{fig9}}
\end{figure}

Until now, we have neglected the effect of grain evaporation at a high
temperature region within the innermost region of an accretion disk.
We take account of grains' thermal evaporation temperature
\citep[\(T=1750\) K for graphite grains, \(T=1400\) K for silicate grains;][]{Hillenbrand92},
and calculate the size of a dead zone for \(Re_{M}=10^3\) cases.
The dead zones now look as in Fig. \ref{fig8}.
We have an inner active region due to thermal collisions of alkali
ions besides the lack of grains in that region.
The corresponding gap-opening masses are plotted in Fig. \ref{fig9}.
Note that in situ planet formation in the inner active region (\(r\le
0.04\) AU) is unlikely.
This is because the region has no dust (therefore core accretion
scenario is impossible) and is gravitationally stable (therefore disk instability
scenario is impossible).
This inner turbulent region is likely to accrete onto the star rather
quickly and form an inner hole due to a higher viscosity.
\section{Dead Zones and Planetary Migration}
In \S1, we argued that the dead zones can slow down planetary migration.
Since the type I migration does not depend on a disk's viscosity, the dead
zones cannot directly affect the low mass planet migration.
However, when a low mass planet starts its migration from outside the
dead zone, it can be strongly affected by the existence of the dead
zone.
This is because of the difference in disk evolution speed between a
dead and an active zone.
Due to the smaller viscosity in a dead zone, the disk mass accreting
toward the central star from an active zone is likely to be accumulated at the
edge of the dead zone \citep{Gammie96}.
If a migrating planet sees this denser region, the inner torque could
be comparable to, or even grater than, the outer torque in magnitude.
This will stall the planet, or even reverse its migration.
As we argued in \S 1, the type I migration of a non-gap-opener can in
principle be divided into two groups: (1) when there is enough mass
accumulation at the edge of a dead zone, a protoplanetary migration is
either stalled or reversed, and (2) when there is not enough mass
accumulation, a planet will migrate into a dead zone and likely to
open a gap.

\begin{figure}
\scalebox{0.3}{
\includegraphics{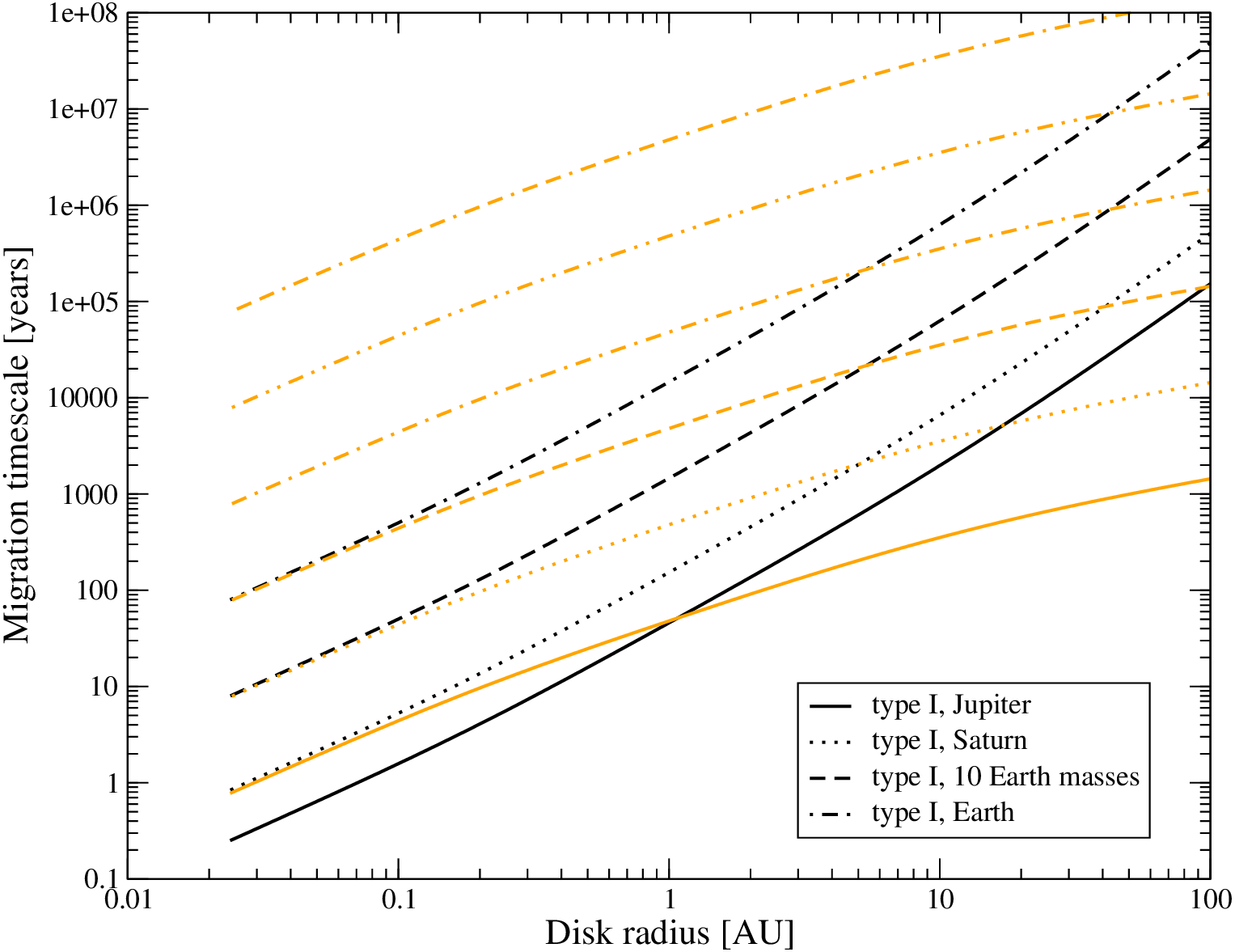}}
\caption[migration]{Migration timescales as a function of disk radius.
Heavy lines show, from bottom to top, type I migration timescales for
Jupiter, Saturn, 10 Earth mass planet, and Earth respectively.  Light
lines show, from bottom to top, type II migration timescales for
\(\alpha_{ss}=1, \ 0.1, \ 0.01, \ 10^{-3}, \ 10^{-4}, \ 10^{-5}\)
respectively. \label{fig10}}
\end{figure}

Roughly speaking, for a planet to enter a dead zone without being
stopped by the mass accumulation at the dead zone edge, it has to migrate
fast enough compared to the disk's mass at around the edge of a dead
zone.
Fig. \ref{fig10} compares type I migration timescales of Jupiter,
Saturn, 10 Earth mass planet, and Earth (heavy lines, from bottom to
top) with disk's viscous evolution timescales (or equivalently type II
migration timescales) for \(\alpha_{ss}=1-10^{-5}\) (light lines, from
bottom to top).
Here we used the following timescale equations \citep{Terquem03ap}
\begin{eqnarray}
\tau_{{\rm I}}  &\sim& 10^{10}\left(\frac{M_{p}}{M_{E}}\right)^{-1}
                \frac{1}{\Sigma \sqrt{r}}\left(\frac{h_{p}}{r_{p}}\right)^2 \\
\tau_{{\rm II}} &=&
\frac{1}{3\alpha_{ss}}\left(\frac{h_{p}}{r_{p}}\right)^{-2}r^{3/2} \ ,
\end{eqnarray}
where \(M_{E}\) is the Earth mass.
For \(\alpha_{ss}=0.01\), it is apparent that Saturn or more massive
planets migrate much faster than disk's mass while 10 Earth mass or
lighter planets migrate slower.
We expect that this mechanism may keep lighter planets beyond a dead
zone and help them to grow larger.

A dead zone's effect on type II migration can be readily seen in Fig. \ref{fig10}.
For example at 10 AU, the migration timescale is \(> 10^4\) years for
\(\alpha_{ss}=0.01\) (in an active zone), and \(> 10^6\) years for
\(\alpha_{ss}=10^{-4}\) (in a dead zone).
Thus, type II planets slow down significantly as soon as they enter
the dead zone due to the lower viscosity there. 
\section{Dead Zones and Planetary Eccentricity}
One of the major surprises about extrasolar planetary systems is the
presence of isolated planets with large orbital eccentricities.
Moreover, all very massive extrasolar planets (\(M_{p}>7M_{J}\)) found
so far have rather large eccentricities (\(e>0.2\)).
We argue that this phenomenon may also be explained by the existence
of a dead zone.

Current models suggest that the growth of planetary eccentricity takes
place while the disk is still present.
In such a situation, eccentricity evolution can occur either
through planet-planet interaction \citep[e.g.][]{Chiang02} or
planet-disk interaction \citep[e.g.][hereafter GS03 and SG04
respectively]{Goldreich03,Sari04}.
Planet-planet interaction would require similar mass (or more
massive) planets to enhance their eccentricities, while all systems
with a very massive planet do not have such a companion.
It is possible that a similar mass companion was ejected out of the
system due to a strong interaction.
The other possibility -- planet-disk interaction -- can increase the
eccentricity if a planet with an initial mild eccentricity opens a
large enough gap (SG04).
Since larger gaps are expected in regions of smaller viscosity, we
suggest that planets caught in gaps within dead zones are likely to
grow their eccentricity.

Conditions during gap formation may be particularly favorable for
eccentricity growth of a planet (GS03, SG04).
Planetary interaction with disks at Lindblad resonances enhances their
eccentricity whereas interaction at corotation resonances damp it.
A necessary condition for eccentricity growth therefore is that the
non-linear saturation of the corotation resonances occur.
This condition implies that the timescale on which the density
gradient is flattened by corotation resonance (\(t_{sat}\)), is
shorter than both the disk viscous timesclale (\(t_{vis}\)) and the
timescale on which the gap of width \(w\) is opened by the principal
Lindblad resonances (\(t_{gap}\)); \(t_{sat}<\left(t_{vis}, \
t_{gap}\right)\) (SG04).
It is also necessary that the eccentricity must grow faster than the
gap grows. 
Slightly modifying their paper, we show that the initial eccentricity
has to be at least
\begin{eqnarray}
\label{eint}
e_{0,{\rm min}}&=&{\rm max}\left[e_{1,{\rm min}}, e_{2,{\rm min}}
\right] \\
e_{1,{\rm min}}&=& \frac{\alpha_{ss}^{2/3}}{M_{p}/M_{*}}
\left(\frac{M_{p}}{M_{d}}\right)^{5/3}
\left(\frac{h_{p}}{r_{p}}\right)^{4/3} \nonumber \\
e_{2,{\rm min}}&=& \left(\frac{M_{p}}{M_{*}}\right)^3 
\left(20\left(\frac{M_{p}}{M_{d}}\right)\right)^{-7} \nonumber 
\end{eqnarray}
for this kind of eccentricity evolution to occur (see Appendix B for
the derivation).

Fig. \ref{fig11e} shows a calculated minimum initial
eccentricity for Jupiter and 10 Jupiter mass planet in the disk of
Fig. \ref{fig8} with \(\alpha_{ss}=0.01\).
It is immediately clear that the required initial eccentricity for
very massive planets is very low (\(e_{0,{\rm min}}\sim10^{-3}\))
inside a dead zone while that outside a dead zone is more than 0.1.
This suggests that eccentricity of heavy planets may have been excited
inside dead zones through disk-planet interaction as they migrated
into them and opened gaps.
It is interesting to note that all high eccentricity (\(e>0.2\))
planets observed so far, including very massive planets
(\(M_{p}>7M_{J}\)), are found between \(0.07 - 10\) AU --
roughly the region of a dead zone.
If a planet's eccentricity is enhanced due to an interaction with a
disk, we expect planets with small eccentricities to be found beyond
10 AU.

The condition that \(e_{1,{\rm min}}=e_{2,{\rm min}}\) in Equation
(\ref{eint}) gives the absolute minimum of the initial eccentricity
required to excite eccentricity growth.
This occurs for roughly a Uranus-mass planet.
We can determine the maximum and minimum mass planets whose initially
required eccentricity is reasonably small (e.g. \(e_{0}=0.01\)).
By setting \(e_{1,{\rm min}}=0.01\), and assuming a disk mass of
\(M_{d}=0.01 M_{\odot}\) as well as a dead zone's viscosity of
\(\alpha_{ss}=5 \times 10^{-5}\), we can determine the maximum mass
planet whose eccentricity is likely to be excited through planet-disk
interaction.
We find that all very massive planets (up to \(\sim 20 M_{J}\)) may
grow their eccentricity inside the dead zone in this way.
Similarly, by setting \(e_{2,{\rm min}}=0.01\), we can determine the
minimum mass planet.
Our calculation shows that planets less than an Earth-mass in a disk of
\(M_{d}=0.01M_{\odot}\) are too light to interact with a disk
and enhance their orbital eccentricity.
Combining these results together, we can see that planets with
masses between Earth and 20 Jupiter masses may enhance their
eccentricity reasonably easily inside the dead zone.
\section{Conclusions}
We calculated the sizes of dead zones and gap-opening masses in models
of stationary protostellar disks.
We performed an extensive parameter search: \(\Sigma_{0}=10^3-10^4 \
{\rm g \ cm^{-2}}\), \(\alpha_{ss}=10^{-3}-1\), \(Re_{M}=10^2-10^4\),
and compared (1) disks in a clustered environment and isolated disks,
and (2) disks with and without cosmic ray ionization. 
Our major findings are as follows:
\begin{enumerate}
\item Dead zones are robust features of protostellar disks.  We took
account of all the major sources of ionization and recombination and
considered other effects like ambipolar diffusion and mass mixing
between active layers and a dead zone. We showed that dead zones
typically stretch out to a few tens of AU.
\item Cosmic ray ionization has a large effect on an ionization
structure at outer part of the disk (beyond 10 AU), but does not
change the sizes of dead zones significantly.  From Fig. \ref{fig2},
it is clear that the ionization of an outer part of the disk is
dominated by cosmic rays.
The sizes of dead zones determined by the total ionization and those
by the X-ray ionization are about the same for the same parameter set
(\(\Sigma_{0}, \ \alpha_{ss}, \ Re_{M}\)), in the same environment
(clustered or isolated star-forming regions), because the ionization
structure in inner dense regions are similar (i.e. dead zone
boundaries intersect with a critical mass ratio at similar radii). 
The difference due to ionization sources become more apparent in a
denser disk (\(\Sigma_{0}\ge10^4 \ {\rm g \ cm^{-2}}\)) because disks
become more optically thick.
\item The size of a dead zone depends mildly on magnetic features of a
disk (\(\alpha_{ss}\) and \(Re_{M}\)), and rather strongly on a disk
surface mass density \(\Sigma_{0}\) (see Fig. \ref{fig5}).  
It is almost independent of disk's environments (clustered or isolated
star-forming regions).
\item Jovian or super Jovian planets are likely to be formed beyond a
dead zone.  Inside dead zones, a gap opens for smaller mass planets --
ice giants or even terrestrial planets (see Fig. \ref{fig4} and \ref{fig5}).   
\item Dead zones may significantly slow down both type I and type II
planet migration (see \S 5).
\item Gaps within the dead zones may be good regions in which planets enhance their orbital
eccentricities via planet-disk interaction. As we discussed in \S 6,
this may be especially true for very massive planets (\(M_{p}>7M_{J}\)).
\end{enumerate}
In our future work, we revisit the problem of planetary migration and
dead zones by using time-dependent calculations.
\section*{Acknowledgments}
We thank Eric Feigelson, David Hollenbach, Doug Lin, Edward Thommes and James Wadsley
for stimulating discussions.
We also thank an anonymous referee, Shu-ichiro Inutsuka, and Jim
Pringle for usueful comments.
S. M. is supported by a SHARCNET Graduate Fellowship while R. E. P. is
supported by a grant from the National Science and Engineering
Research Council of Canada (NSERC).
\bibliography{REF}
\bibliographystyle{mn2e}
\clearpage
\clearpage
\clearpage
\clearpage
\clearpage
\clearpage
\clearpage
\clearpage
\begin{figure}
\scalebox{0.3}{
\includegraphics{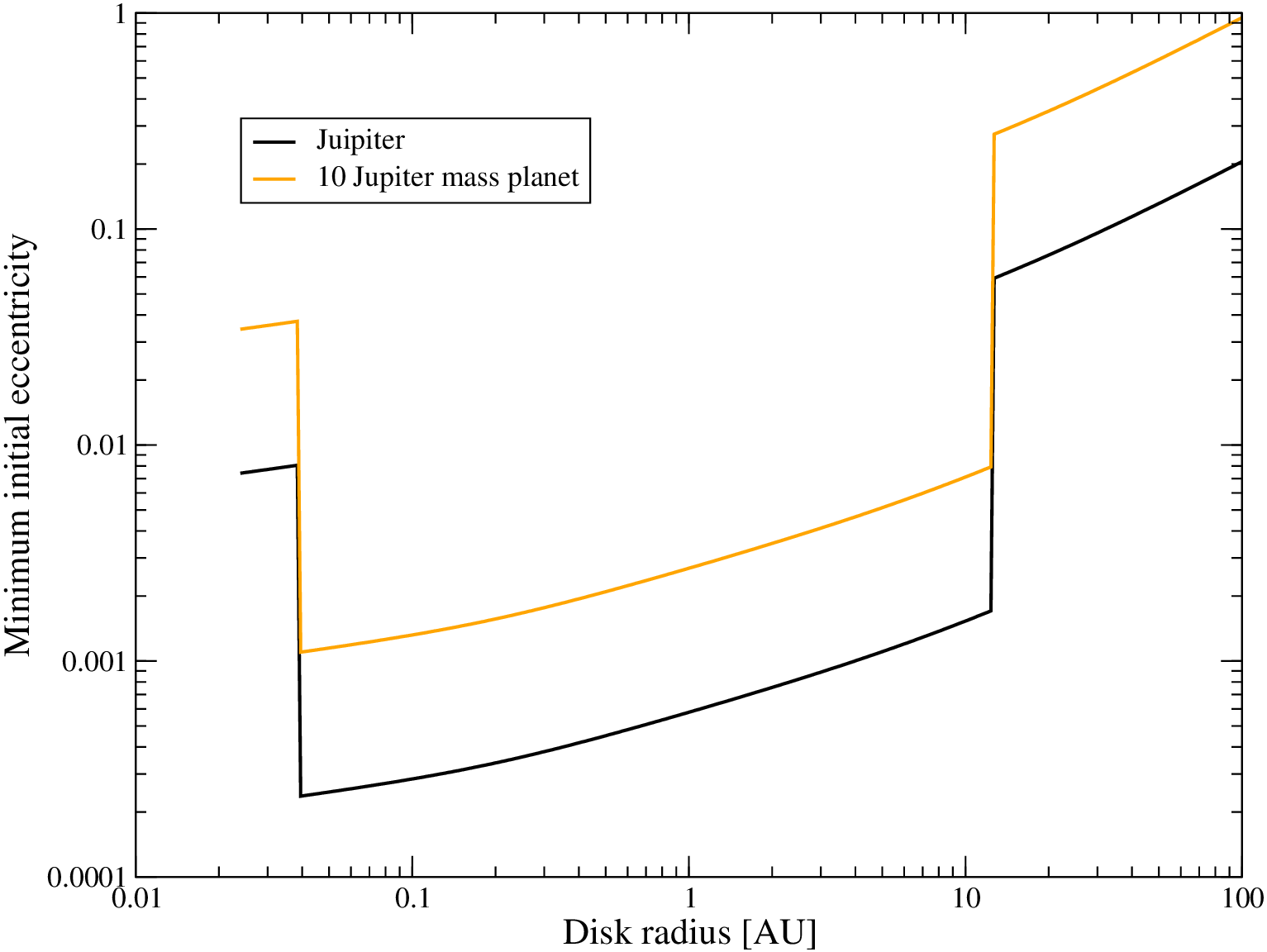}}
\caption[eccentricity]{The minimum initial eccentricity required for
eccentricity evolution via planet-disk interaction
\citep{Goldreich03}.  Heavy line is for a Jupiter mass planet and light
line is for a 10 Jupiter mass planet.  Obviously, eccentricity growth
is more likely for a planet in the dead zone. \label{fig11e}}
\end{figure}
\clearpage
\appendix
\begin{table}
\caption[recombination]{Recombination coefficients in rate equations. \label{tb1}}
\begin{center}
\begin{tabular}{|l|l|l|} \hline
symbols & recombination rate coefficients [\({\rm cm^3 s^{-1}}\)] &  pairs \\ \hline \hline
\(\beta\) & \(3 \times 10^{-6} \ T^{-1/2}\) & electrons -- molecular ions \\ \hline
\(\beta_{r}\) & \(3 \times 10^{-11} \ T^{-1/2}\) & electrons -- metal ions \\ \hline
\(\beta_{t}\) & \(3 \times 10^{-9}\) & molecular ions -- metal atoms \\ \hline
\(\beta_{1}\) & \(\pi a^2 \sqrt{\frac{8k_{B}T}{\pi m_{e}}}0.6\) & electrons -- neutral grains \\ \hline
\(\beta_{2}\) & \(\pi a^2 \sqrt{\frac{8k_{B}T}{\pi m_{e}}}(1-\frac{q^2}{a k_{B} T})\) & electrons -- positively charged grains \\ \hline
\(\beta_{3}\) & \(\pi (2a)^2 \sqrt{\frac{8k_{B}T}{\pi m_{G}}}(1-\frac{q^2}{2a k_{B} T})\) & positively charged grains -- negatively charged grains \\ \hline
\(\beta_{4}\) & \(\pi a^2 \sqrt{\frac{8k_{B}T}{\pi m_{m}}}\) & molecular ions -- neutral grains \\ \hline
\(\beta_{5}\) & \(\pi a^2 \sqrt{\frac{8k_{B}T}{\pi m_{m}}}(1-\frac{q^2}{a k_{B} T})\) & molecular ions -- negatively charged grains \\ \hline
\(\beta_{6}\) & \(\pi a^2 \sqrt{\frac{8k_{B}T}{\pi m_{M}}}\) & metal ions -- neutral grains \\ \hline
\(\beta_{7}\) & \(\pi a^2 \sqrt{\frac{8k_{B}T}{\pi m_{M}}}(1-\frac{q^2}{a k_{B} T})\) & metal ions -- negatively charged grains \\ \hline
\end{tabular}
\end{center}
\end{table}
\clearpage
\section{Effect of Grains on Disk Ionization Balance}
To calculate an electron fraction by including various recombination sources, we write the rate equations for electrons, molecular ions, neutral and singly charged grains as follows:
\begin{eqnarray}
\frac{dn_{e}}{dt}&=&\xi n_{n} -\beta \ n_{m^+}n_{e} -\beta_{r} \ n_{M^+}n_{e} -\beta_{1} \ n_{G}n{e} \nonumber \\ 
                 & & -\beta_{2} \ n_{G^+}n_{e} \\
\frac{dn_{m+}}{dt}&=&\xi n_{n} -\beta \ n_{e}n_{m+} -\beta_{t} \ n_{M}n_{m+} -\beta_{4} \ n_{G}n_{m+} \nonumber \\ 
                  & & -\beta_{5} \ n_{G^-}n_{m+} \\
\frac{dn_{G}}{dt}&=&\beta_{2} \ n_{G^+}n_{e} +\beta_{5} \ n_{G^-}n_{m^+} +\beta_{7} \ n_{G^-}n_{M^+} \nonumber \\ 
                 & & + 2 \ \beta_{3} \ n_{G^+}n_{G^-} -\beta_{1} \ n_{e}n_{G} -\beta_{4} \ n_{m^+}n_{G} \nonumber \\
                 & & -\beta_{6} \ n_{M+}n_{G} \\
\frac{dn_{G^+}}{dt}&=&\beta_{4} \ n_{G}n_{m^+} +\beta_{6} \ n_{G}n_{M^+} -\beta_{2} \ n_{e}n_{G^+} \nonumber \\ 
                   & & -\beta_{3} \ n_{G^-}n_{G^+} \\
\frac{dn_{G^-}}{dt}&=&\beta_{1} \ n_{G}n_{e} -\beta_{3} \ n_{G^+}n_{G^-} -\beta_{5} \ n_{m^+}n_{G^-} \nonumber \\ 
                   & & -\beta_{7} \ n_{M+}n_{G^-} \\
n_{e}&=&n_{m^+} +n_{M+} +n_{G^+} -n_{G^-} \ .
\end{eqnarray}
Here, \(n_{e}, \ n_{m}, \ n_{M}, \ n_{G}\), and \(n_{n}\) are the
density of electrons, molecules, metals, grains, and neutral
hydrogens, while the superscript of \(+\) or \(-\) indicate the
elements are positively or negatively charged respectively.
Also, \(\xi\) is the total ionization rate \(\beta\)s are
recombination rate coefficients (see Table \ref{tb1} for details).
We solve these differential equations together with charge
conservation by using a semi-implicit extrapolation method (a generalized
Bulirsch-Stoer method) \citep[stifbs.f90 in][]{NRs}.
This calculation takes about a few hours to run on the average workstation.

Table \ref{tb1} shows all recombination coefficients
appeared in above equations, where \(k_{B}\) is the Boltzmann
constant, \(T\) is the disk temperature, \(a\) is a grain radius, and
\(q\) is the charge.
Also, \(m_{e}, \ m_{m}, \ m_{M}\), and \(m_{G}\) are the mass of
electrons, molecules, metals, and grains respectively. 
The recombination rate coefficients \(\beta, \ \beta_{r}, \
\beta_{t}\) are taken from \cite{Fromang02}, and the rest are obtained
by following the method by \cite{Sano00}.
\section{Minimum initial eccentricities}
In \S 5, we argued that there is the minimum initial eccentricity
required to enhance a planetary eccentricity via planet-disk
interaction.
\cite{Sari04} suggested that for the eccentricity to grow, (1) the
corotation resonances must saturate, and (2) eccentricity must grow
faster than a gap opens.
Also, they claimed that eccentricity does not significantly decay as
long as (3) corotation resonances are more than 5 \% saturated.

For the first point, two equations must be satisfied: \(t_{sat}<
t_{vis}\); which may be written as
\begin{equation}
e > \frac{\alpha_{ss}^{2/3}}{M_{p}/M_{*}}
\left(\frac{w}{r_{p}}\right)^{5/3}
\left(\frac{h_{p}}{r_{p}}\right)^{4/3} \equiv e_{1}
\end{equation}
as well as \(t_{sat}< t_{gap}\); which becomes
\begin{equation}
e > \left(\frac{M_{p}}{M_{*}}\right)^3
\left(\frac{r_{p}}{w}\right)^{7} \equiv e_{2}
\end{equation}
where \(w\) is a gap width and \(M_{*}\) is the stellar mass (SG04). 

Combining the second and third points, we can write 
\begin{equation}
\frac{M_{p}}{M_{d}}<\frac{w}{r_{p}}<20\frac{M_{p}}{M_{d}} \ ,
\end{equation}
where \(M_{d}\) is a disk mass.
The left inequality arises because a gap has to open slower than the
eccentricity evolution timescale, while the right inequality comes in
so that the significant decay won't happen.

Writing these as the minimum and maximum ratios of a gap width and a
disk radius, we can rewrite first two conditions as follows:
\begin{eqnarray}
e &>& e_{1} \\ \nonumber
  &>&  \frac{\alpha_{ss}^{2/3}}{M_{p}/M_{*}}
\left(\frac{w}{r_{p}}\right)^{5/3}_{{\rm min}}
\left(\frac{h_{p}}{r_{p}}\right)^{4/3} \equiv e_{1,{\rm min}} \\ 
e &>& e_{2} \\ \nonumber
  &>& \left(\frac{M_{p}}{M_{*}}\right)^3
\left(\frac{w}{r_{p}}\right)^{-7}_{{\rm max}} \equiv e_{2,{\rm min}} \ .
\end{eqnarray}

Thus, the eccentricity must be at least \(e>{\rm max}\left[e_{1,{\rm
min}},e_{2,{\rm min}}\right]\) to
be excited, which is Equation (\ref{eint}).
\label{lastpage}
\end{document}